\documentclass[10pt]{article}
\usepackage[utf8x]{inputenc}
\pdfoutput=1
\usepackage{amssymb,amsmath,mathrsfs,enumerate}
\usepackage{mwe}
\usepackage{graphicx}
\usepackage{siunitx}
\usepackage{rotate,multicol}
\usepackage{float}
\usepackage[subfigure]{tocloft}
\usepackage[caption=false]{subfig}
\usepackage[margin=10pt,labelfont=bf]{caption}
\usepackage{cite}
\usepackage{soul}
\usepackage{epstopdf}
\usepackage{epsfig}
\usepackage{gensymb}
\usepackage{svg}
\usepackage[colorlinks=true,
            linkcolor=blue,
            urlcolor=blue,
            citecolor=teal]{hyperref}
\usepackage{multirow}
\usepackage{siunitx}
\usepackage{wrapfig,braket}
\usepackage{lipsum}
\newcommand{\blds}[1]{\mbox{\small  $#1$}}
\usepackage[normalem]{ulem}
\newcommand{\stkout}[1]{\ifmmode\text{\sout{\ensuremath{#1}}}\else\sout{#1}\fi}

\makeatletter
\let\Hy@linktoc\Hy@linktoc@page
\makeatother

\usepackage{color}
\definecolor{ourcolor}{rgb}{0.7, 0.25, 0.05}
\usepackage{tikz,braket}
\long\def\rpl#1!!#2!!{\textcolor{red}{#1} \textcolor{blue}{#2}}

\let\hat=\widehat

\def \order(#1){{\mathcal O} \left(#1 \right)}

\evensidemargin=\oddsidemargin
\newcommand{\be}{\begin{equation}}
\newcommand{\ee}{\end{equation}}
\makeatletter
\renewcommand{\fnum@table}{\textbf{\tablename~\thetable}}
\renewcommand{\fnum@figure}{\textbf{\figurename~\thefigure}}
\makeatother
\def\wt{\widetilde}
\def\({\left(}
\def\){\right)}
\def\[{\left[}
\def\]{\right]}

\newcommand{\ba}{\begin{array}}
\newcommand{\ea}{\end{array}}
\newcommand{\bd}{\begin{displaymath}}
\newcommand{\ed}{\end{displaymath}}
\makeatletter
\newcommand*{\rom}[1]{\expandafter\@slowromancap\romannumeral #1@}
\makeatother
\def\bt{\begin{table}}
\def\et{\end{table}}
\def\bc{\begin{center}}
\def\ec{\end{center}}
\def\bi{\begin{itemize}}
\def\ei{\end{itemize}}
\def\bw{\begin{widetext}}
\def\ew{\end{widetext}}

\def\bea{\begin{eqnarray}}
\def\eea{\end{eqnarray}}
\def\beas{\begin{eqnarray*}}
\def\eeas{\end{eqnarray*}}

\def\N0{\widetilde{\chi}^0}

\def\a{\alpha}

\def\b{\beta}

\def\G{\Gamma}

\def\l{\lambda}

\catcode`@=12

\allowdisplaybreaks

\title{\color{black}{\bf Zooming in on eV-MeV Scale Sterile Neutrinos in light of Neutrinoless Double Beta Decay}}

\author {\bf Tapoja Jha$^{a,b,c}$\footnote{tapoja.phy@gmail.com} 
\hspace{4pt} Sarif Khan$^{d,}$\footnote{sarif.khan@uni-goettingen.de} 
\hspace{4pt} Manimala Mitra$^{a,b}$\footnote{manimala@iopb.res.in}
\hspace{4pt} Ayon Patra$^{e,}$\footnote{ayon@okstate.edu}
\\[10pt]
\small\em $^a$Institute of Physics, Sachivalaya Marg, Bhubaneswar, Odisha 751005, India\\
\small\em $^b$Homi Bhabha National Institute, Training School Complex, Anushakti Nagar, Mumbai 400085, India\\
\small\em $^c$School of Physical Sciences, Indian Association for the Cultivation of Science, \\ \small\em 2A \& 2B Raja S.C. Mullick Road, Kolkata-700 032, India\\
\small\em $^d$Instit\"ut f\"ur Theoretische Physik, Georg-August-Universit\"at 
G\"ottingen,\\
\small\em Friedrich-Hund-Platz 1, 37077 G\"ottingen, Germany\\
\small\em $^e$Division of Physics, School of Advanced Sciences, VIT University, Chennai Campus, Chennai 600127, India
}

\date{}

\begin{document}

\maketitle

\begin{abstract}
The existence of light sterile neutrinos, as predicted in several models, can help to explain a number of observations starting from dark mater to recent anomalies in short baseline experiments. In this paper we consider two models - Left-Right Symmetric Zee model and Extended Seesaw model, that can naturally accommodate the presence of light sterile neutrinos in the eV to MeV mass scale. We perform a detailed study on the neutrinoless double beta decay process which receives major contributions from diagrams involving these light sterile neutrinos. Considering a number of theoretical and experimental constraints, including light neutrino masses and mixings, unitarity of the mixing matrix etc., we compare our predicted values of the half-life of neutrinoless double beta decay with the experimental limits. This can put significant constraints on the neutrino mass, active-sterile neutrino mixing and several other important parameters in these models.
\end{abstract}

\textcolor{blue}{{\bf Keywords:} Sterile neutrino; Neutrinoless double beta decay; Left-right symmetry; Extended Seesaw.}

\newpage

\hrule \hrule
\tableofcontents
\vskip 20pt
\hrule \hrule 

%%%%%%%%%%%%%%%%%%%%%%%%%%%%%%%%%%%%%%%%%%%%%%%%%%%%
%%%%%%%%%%%%%%%%%%%%%%%%%%%%%%%%%%%%%%%%%%%%%%%%%%%%
%%%%%%%%%%%%%%%%%%%%%%%%%%%%%%%%%%%%%%%%%%%%%%%%%%%%
%%%%%%%%%%%%%%%%%%%%%%%%%%%%%%%%%%%%%%%%%%%%%%%%%%%%

\section{Introduction}
\label{s:Intro}
%%%%%%%%%%%%%%%%%%%%%%%%%%%%%%%%%%%%%%%%%%%%%%%%%%%%%%%%%%%%%%%%%%%%%%%%%%%%%%%%%%%%%%%%%%%%%%%%%%%%%%%%%%%%%%%%%%%%%%%%%%%%%%

%The observation of neutrino oscillation, existence of dark matter and the baryon asymmetry of the universe are some of the developments that have ushered us into a new era and compelled us to think beyond the known particle physics models. 
The Standard Model (SM) of particle physics, despite its major successes, is unable to explain the observed light neutrino mass splittings and their mixings, which provides a strong motivation to invoke beyond the Standard Model (BSM) physics. The two observed neutrino mass splittings are $\Delta m^2_{12}\sim 10^{-5}$ $eV^2$,  $|\Delta m^2_{13}| \sim 10^{-3}\, \textrm{eV}^2$ while the best-fit values of the neutrino mixing angles are $\theta_{12}\sim 34^\circ, \theta_{23} \sim 48^\circ$ and $\theta_{13}\sim 8^{\circ}$~\cite{Esteban:2020cvm}.  Though neutrinos are massless in SM, a number of BSM theories have been proposed that successfully explain neutrino masses and mixings. One of the most appealing frameworks
to generate Majorana masses of  light neutrinos  is via seesaw, where the dimension-5 lepton number violating operator generates the mass term after electroweak symmetry breaking \cite{Minkowski:1977sc,Sawada:1979dis,Weinberg:1979sa,Wilczek:1979hc,Glashow:1979nm,GellMann:1980vs,Mohapatra:1979ia}. The type-I seesaw serves as  the most economical framework, as the model in addition to the SM particles is minimally extended by gauge singlet right-handed neutrinos. Another popular class of mechanism   is the radiative mass generation \cite{Zee:1985id,Babu:1988ki,Ma:1998dn,Bonnet:2012kz,Sierra:2014rxa}, where neutrino mass is generated via loop effect.  In this work, we have considered a variation of the type-I seesaw model referred as  Extended Seesaw model \cite{Kang:2006sn,Mitra:2011qr} and a left-right symmetric extension  of radiative neutrino mass model \cite{Mohapatra:1974hk,Senjanovic:1975rk,FileviezPerez:2017zwm,Khan:2018jge}.

The type-I seesaw model is the most economical, as the SM particle content is expanded with at least two heavy gauge singlet right-handed neutrinos which participate in   light neutrino mass generation via seesaw mechanism. However, the drawback of this simplest model is that the mixing of these right-handed neutrinos with SM neutrinos are tightly constrained by eV light neutrino mass constraint,  making the detection prospect of these right-handed neutrinos at experiments challenging.  In Extended Seesaw, as the name suggests, more singlet neutrinos with large mixings are introduced with the possibility that some of them remain light and can be detected  in experiments. The other popular mechanism for neutrino mass generation is through loop-induced processes. One of the simplest examples of this process is realized in the Zee model where the introduction of a doublet scalar and a charged singlet scalar can generate  neutrino masses at the one-loop level. Though the simplest form of Zee model~\cite{Zee:1980ai} cannot satisfy neutrino oscillation data~\cite{Frampton:2001eu,Koide:2002uk,He:2003ih}, its left-right symmetric extension however is consistent with experimental observations~\cite{FileviezPerez:2017zwm,FileviezPerez:2016erl}. Here the Majorana mass of the left-handed and right-handed neutrinos are generated at the one-loop level while the Dirac mass term arises at the tree level from the Yukawa interactions. Finally the light neutrino masses are obtained by a type-1 seesaw mechanism but with the exception that the right-handed neutrinos can be light and hence offer better detection prospects.

Both of the above mentioned  models  can  accommodate light right-handed  neutrinos with masses ranging in the  eV to MeV scale. An eV scale sterile neutrino %\footnote{Although having eV scale sterile neutrino in the model is highly constrained from cosmology \cite{Hamann:2011ge}, there are also many counter-studies on how to evade the cosmological bounds 
%\cite{Kim:2018uht,Pires:2019elj}.}
 is well motivated, as this can explain LSND anomaly \cite{Karagiorgi:2012usa, Antonello:2012pq, Aguilar:2001ty}.
Recently this anomaly has also been favoured by the MiniBooNE collaboration \cite{AguilarArevalo:2008rc} but at the same time the data has been disfavoured by the KARMEN \cite{Armbruster:2002mp} and MINOS \cite{MINOS:2016viw} observations. These issues may be finally tackled by the upcoming DUNE experiment \cite{DUNE:2018tke,DUNE:2020ypp}. Further hints regarding the presence of an eV scale sterile neutrino comes from the reactor anti-neutrino anomaly (RAA)~\cite{Mention:2011rk,Abazajian:2012ys} and the Gallium anomaly \cite{Kaether:2010ag,Abdurashitov:2009tn}. A keV scale sterile neutrino can be an excellent candidate for warm dark matter. Several disagreements between the cosmological observations and the N-body simulations of structure formations can be solved by introducing a keV scale warm dark matter candidate \cite{Adhikari:2016bei}. The presence of an MeV sterile neutrino, on the other hand, can produce several observable astrophysical signals, such as, its effect on the Cosmic Microwave Background spectrum \cite{Sabti:2019mhn} and by producing X-ray photons which may be observable in satellite based X-ray experiments. 

If the right-handed neutrinos are Majorana particles, they can give rise to additional contributions to the neutrinoless double beta decay ($0\nu\beta\beta$) process. The $0\nu\beta\beta$ process is the  transition $(A,Z) \rightarrow (A, Z+2)+2e^-$ with no neutrino being emitted~\cite{Majorana:1937vz,Racah:1937qq,Furry:1939qr}. The process is lepton number violating~\cite{Vergados:2016hso,DellOro:2016tmg}. Depending on the mixing of the right-handed neutrinos with active neutrinos in type-I/Extended Seesaw, or the interaction of these right-handed neutrinos with the right-handed gauge  boson in left-right symmetric extension, these right-handed neutrino states may give significant contributions in $0\nu\beta\beta$ process compared to the three SM neutrino contributions, and thus  opening up the scope of detection of these Majorana neutrinos via $0\nu \beta \beta$ process. A number of experiments have searched for this process and the non-observation of signal has given bound on the half-life $T_{1/2}^{0\nu}$ of $0\nu \beta \beta$~\cite{Agostini:2018tnm,KamLAND-Zen:2016pfg,Arnaboldi:2002du,Arnold:2010tu}. The limit obtained  on $T_{1/2}^{0\nu}$ for $^{76}{\rm Ge}$ is  $T_{1/2}^{0\nu} > 8.0 \times 10^{25}$ year from  GERDA-II \cite{Agostini:2018tnm}; whereas, at $90\%$ C.L. the KamLAND-Zen experiment has set a more stringent lower limit on the half-life of $^{136}{\rm Xe}$ isotope as $T_{1/2}^{0\nu} > 1.07 \times 10^{26}$ year~\cite{KamLAND-Zen:2016pfg}.

In the present work, we consider Left-Right Symmetric Zee model (LRS Zee) and Extended Seesaw model that naturally accommodate light scale sterile neutrinos with masses $\sim$ eV to MeV. Our main goal is to study the $0\nu\beta\beta$ phenomenology for these two models. The left-right symmetric extension of the Zee model \cite{FileviezPerez:2016erl,FileviezPerez:2017zwm} presents a unique scenario where the model can be tested at the collider experiments as well as the neutrino experiments. It may have observable signals at the hadron \cite{FileviezPerez:2016erl,FileviezPerez:2017zwm} and lepton colliders \cite{Khan:2018jge} and most notably can be accessed at the very early stage run of the upcoming $e^{+} e^{-}$ colliders. In addition to satisfying all the neutrino mass and mixing constraints, this model can also give rise to several new $0\nu\beta\beta$ process which can significantly enhance the decay rate. This results in a marked decrease in the half-life of the $0\nu\beta\beta$ decay process in this model. We study the variation of $T_{1/2}^{0\nu}$ with respect to different model parameters and identify three of them which are most significant. These three parameters are the lightest neutrino mass, the Dirac CP phase and the mixing angle between the left and right gauge bosons. By varying these parameters we identify the regions which can be ruled out from the experimental limits on the half-life of $0\nu\beta\beta$ process. In the case of extended seesaw mechanism, we first give an approximate analysis considering only one generation right-handed neutrino and one generation active neutrino. Subsequently, we present a realistic analysis of the half-life with three generation  active neutrinos and  six generation  right-handed neutrinos. We have considered all the constraints arising  both from theory and experiments. For the active neutrinos, we have considered bounds on the mass square differences, three mixing angles in agreement with neutrino oscillation data~\cite{Esteban:2020cvm} and the limit on the sum of the masses of active neutrinos which comes from Planck satelite experiment~\cite{Ade:2015xua}. We have ensured a mass hierarchy among these active and right-handed neutrinos to validate the seesaw approximation for this model, as well as have considered constraints from non-unitarity~\cite{Blennow:2016jkn}. We have calculated the $0\nu\beta\beta$ decay contribution considering all the required model parameters which pass all the aforementioned constraints and have checked if the predicted contribution satisfies the corresponding experimental limits~\cite{KamLAND-Zen:2016pfg}. 

The paper is organised in the following way. In Section~\ref{s:LRZee} we present a detailed study of the $0\nu \beta \beta$ process in the LRS Zee model. This is followed by Section~\ref{s:excsaw}, where we give a detailed description of extended seesaw scenario and the analysis of the model with respect to many theoretical and experimental aspects, {\em e.g.}, neutrinuo oscillation data, $0\nu\beta\beta$ decay, unitarity and others. Finally, we present our conclusions in Section~\ref{s:conclusion}.

\section{Left-Right Symmetric Zee Model and Analysis}
\label{s:LRZee}
The Zee model~\cite{Zee:1980ai} is one of the simplest extensions of the Standard Model (SM) which can explain the origin of neutrino mass. By extending the SM framework with an extra doublet and a charged singlet scalar, neutrino masses can be successfully generated at the one-loop level. This simplest form of the Zee model, though, is found to be incompatible with the neutrino experimental data~\cite{Frampton:2001eu,Koide:2002uk,He:2003ih,Herrero-Garcia:2017xdu} and one needs to extend it further in order to get a viable scenario to explain all the neutrino oscillation constraints. The left-right symmetric extension of the Zee model~\cite{FileviezPerez:2016erl,FileviezPerez:2017zwm} provides an alternate model framework which can easily explain the neutrino oscillation data as well as provide interesting flavor violating signals and unique collider signatures. 

The pair production and decay of the singly charged Higgs boson can produce final states with two charged leptons (with either same or different flavors) and missing transverse energy. This process has been studied in the context of the Large Hadron Collider (LHC) in Ref.~\cite{FileviezPerez:2016erl,FileviezPerez:2017zwm} and for the International Linear Collider (ILC) and the Compact Linear Collider (CLIC) in Ref.~\cite{Khan:2018jge}. The charged singlet scalar pair-production cross-section at hadron colliders is quite small and is dominated by the photon mediated process, while it can become significantly larger in the electron-positron collider due to the right-handed neutrino mediated $t$-channel  diagram. Thus the ILC and CLIC experiments may be able to observe such a particle with very low integrated luminosity of only 1-3 $\text{fb}^{-1}$, see \cite{Khan:2018jge} for details. Below we present a brief discussion on the model.

\subsection{Model}
\label{s:LRZee_model}

Other than the three SM neutrinos, the neutrino sector in this model also contains light right-handed neutrinos with masses ranging from an eV to around 100 eV. Their masses depend  on the coupling of right-handed neutrinos with the charged singlet scalar. The Yukawa Lagrangian in this model is given as \cite{FileviezPerez:2016erl,FileviezPerez:2017zwm,Khan:2018jge}
\begin{eqnarray}
\mathcal{L}_Y&=&Y_{ij}^q\overline{Q}_{Li}\Phi Q_{Rj}+\widetilde{Y}_{ij}^q\overline{Q}_{Li}\widetilde{\Phi}Q_{Rj}+Y_{ij}^l\overline{l}_{Li}\Phi l_{Rj}+\widetilde{Y}_{ij}^l\overline{l}_{Li}\widetilde{\Phi}l_{Rj} \notag \\
&+&\l_{L_{ij}} l^T_{Li} i \tau_2 l_{Lj} \delta^+ + \l_{R_{ij}} l^T_{Ri} i \tau_2 l_{Rj} \delta^+  + H.C.~~~~
\end{eqnarray}
where $\Phi$ and $\delta$ are the bidoublet and charged singlet scalars with $Y$ and $\lambda$ being their respective Yukawa couplings. The Majorana masses of all the neutrinos are generated at the one-loop level and as a result, they remain quite light. In order to understand the neutrino sector, we also need to  understand the scalar sector of the model. 

{\bf \it Charged scalar spectrum:} The minimal Higgs sector in this model consists of a bidoublet, two doublets and a charged singlet field given as

\begin{eqnarray}
H_R(1,1,2,1)&=&\left (\begin{array}{c}
H_R^+ \\H_R^0 \end{array} \right ),~~
H_L(1,2,1,1)=\left (\begin{array}{c}
H_L^+ \\H_L^0 \end{array} \right ),~~
\Phi(1,2,2,0)={\left (\begin{array}{cc}
\phi^{0}_1 & \phi^{+}_{2} \\ \phi^{-}_{1} & \phi^{0}_{2} \end{array} \right )},~~ \delta(1,1,1,2) = \delta^+ .~~~~~
\end{eqnarray}
The $SU(2)_R \times U(1)_{B-L}$ symmetry is broken down to  $U(1)_Y$ as the neutral component of the right-handed doublet $H_R$ acquires a non-zero vacuum expectation value (VEV). The bidoublet $\Phi$ is required to generate the quark and charged lepton masses and Cabibbo-Kobayashi-Maskawa (CKM) mixing angles. The neutrino masses, similar to the Zee model, are generated at the one-loop level due to the presence of the charged singlet scalar field. The charged Higgs bosons play an important role in the generation of the one-loop neutrino masses as their masses and mixings become important parameters in the expression for the induced neutrino Majorana masses. So before we study the neutrino mass generation mechanism it is important to define the mass basis for the charged Higgs bosons. There are in total five charged Higgs states which mix to give five mass eigenstates through the rotation
\begin{equation}
\begin{pmatrix}
{\phi_1^-}^*\\ \phi_2^+ \\ H_R^+ \\H_L^+ \\ \delta^+
\end{pmatrix} = V \begin{pmatrix}
H^+_1 \\  H^+_2 \\H^+_3 \\G_1^+ \\G_2^+
\end{pmatrix},
\label{eq:hmix}
\end{equation}
where $V$ is the $5 \times 5$ charged Higgs mixing matrix. There are three physical Higgs bosons $H_1^+, H_2^+, H_3^+$ which will contribute to the neutrino masses and two Goldstone states $G_1^+, G_2^+$ which are eaten up by the $W_R$ and $W$ boson as their longitudinal degrees of freedom. In this model, the ${\phi_1^-}^*$ and the $H_R^+$ become the Goldstone bosons while the other three charged states mix to form the physical Higgs bosons. Constraints from flavor violating process further require the mass of the second bidoublet scalar $\phi_2^+$ to be heavier than around 15 TeV. Thus the largest off-diagonal contributions to $V$ comes from the mixing of the charged singlet $\delta^+$ with the left-handed charged Higgs boson $H_L^+$. These elements will play an important role in the neutrino mass generation as well.

\begin{center}
\begin{figure}
~~~~~~~~~~~~~~~~~~~~~~~~~~~~~~~~\includegraphics[scale=0.3]{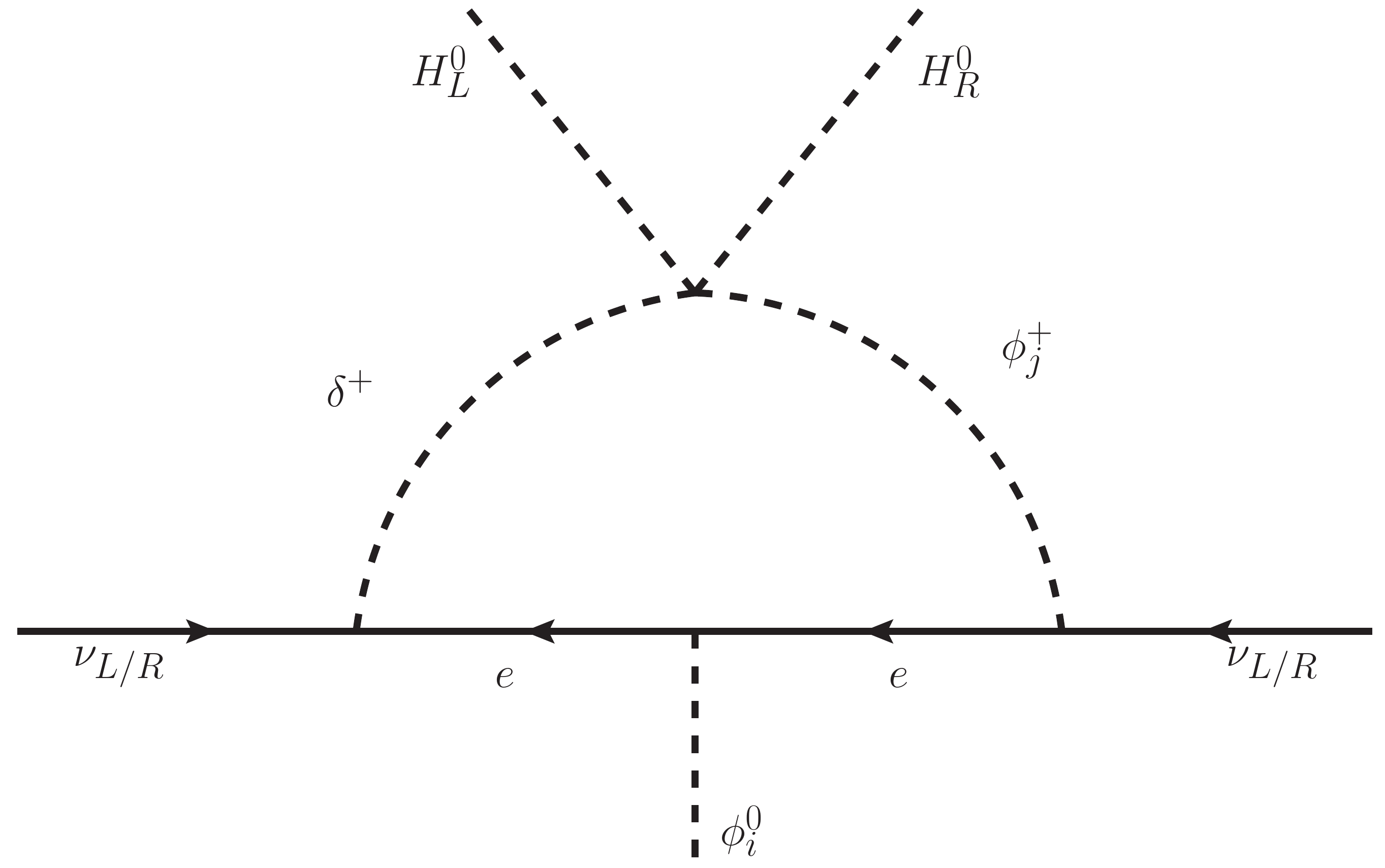}
\caption{Neutrino Majorana mass generation at one-loop in the LRS Zee model.}
\label{fig:nmass}
\end{figure}
\end{center}

{\bf \it Neutrino mass and mixings:} The neutrino sector consists of three left-handed and three right-handed neutrinos. The absence of triplet scalars in the model prevents us from writing a Majorana mass term for the neutrinos. All the neutrino Majorana masses here are generated at the one-loop level and hence are quite small. The lightest right-handed neutrino mass ranges from a few eV to a few hundred eV and the other right-handed neutrinos also remain lighter than an MeV. The Dirac masses are thus required to be quite small as well, so as to satisfy the experimentally observed neutrino masses and mixings\footnote{Detailed analysis of the neutrino sector of LRS Zee model is presented in Ref.~\cite{Khan:2018jge}}. The one-loop Feynman diagram for the generation of neutrino Majorana masses is given in Fig.~\ref{fig:nmass}. The corresponding expressions for the neutrino Majorana masses in this case are given as~\cite{FileviezPerez:2016erl}:
\begin{eqnarray}
\label{eq:RHM}
\displaystyle {(M_\nu^L)}^{\alpha \gamma}&&=\frac{1}{4\pi^2}\lambda'_L{^{\alpha \beta}}m_{e_\beta}\sum_{i=1}^3 \text{Log}\left(\frac{M_{h_i}^2}{m_{e_\beta}^2}\right) \times V_{5i}\left [ (Y_l^\dagger)^{\beta \gamma}V_{2i}^*-(\wt Y_l^\dagger)^{\beta \gamma}V_{1i}^* \right ] \ + \ \alpha \leftrightarrow \gamma 
\,,\notag \\
\displaystyle {(M^R_\nu)}^{\alpha \gamma}&&=\frac{1}{4\pi^2}\lambda'_R{^{\alpha \beta}}m_{e_\beta}\sum_{i=1}^3\text{Log}\left(\frac{M_{h_i}^2}{m_{e_\beta}^2}\right) \times V_{5i}\left[(Y_l)^{\beta \gamma}V_{1i}^*-(\wt Y_l)^{\beta \gamma}V_{2i}^*\right] \ + \ \alpha \leftrightarrow \gamma \,. 
\end{eqnarray}
In the above, $\l'_{L/R}{^{\a \b}} = \l_{L/R}^{\a \b} - \l_{L/R}^{\b \a}$, $m_{e_\beta}$ and $M_{h_i}$ are the charged lepton and Higgs boson masses and $V_{ij}$ are the charged Higgs boson mixings given in Eq.~\ref{eq:hmix}. For our calculations we will consider the case with $\l'_L$ = 0 as was discussed in \cite{Khan:2018jge}. This will thus give us a $6 \times 6$ neutrino mass matrix given as
\begin{equation}
M_\nu = \begin{pmatrix}
0&M_D\\M_D^T&M_R
\end{pmatrix},
\end{equation}
where $M_D = Y^l <\phi_1^0> +~ \widetilde{Y^l} <\phi_2^0>$. We thus have a scenario that is very similar to the type-I seesaw mechanism, {\it i.e.,} the light and heavy neutrino mass matrix after persuing a block-diagonalization becomes,
\begin{equation}
\mathcal{M}_{\nu}=-M_D M^{-1}_R M^T_D, ~~~ \mathcal{M}_n=M_R.
\label{eq:lrzeeblock}
\end{equation}
The neutrino rotation matrix, taking it from flavor to mass eigenstates, will be a $6\times6$ matrix which we can write as
\begin{equation}
\mathcal{V}=\begin{pmatrix}
\mathbb{U}&\mathbb{S}\\\mathbb{T}&\mathbb{V}
\end{pmatrix},
\label{eq:numix}
\end{equation} 
such that 
\begin{equation}
\mathcal{V}^T M_\nu \mathcal{V} = \begin{pmatrix}
\hat{M}_\nu&0\\0&\hat{M}_N
\end{pmatrix}.
\end{equation}
Here $\hat{M}_\nu={\text {diag}}(m_1,m_2,m_3)$ and $\hat{M}_N={\text {diag}}(M_1,M_2,M_3)$ are diagonal matrices consisting of the light and heavy neutrino masses,  respectively.

\begin{center}
\begin{figure}[h!]
~~~~~~~~~~~~~~~~~~~~~~~~~~~~~~~~~~~~~~~~~~\includegraphics[scale=0.45]{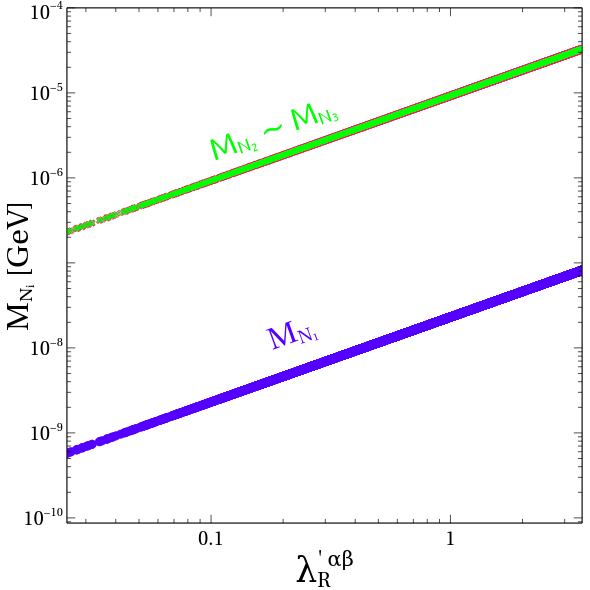}
\caption{Right-handed neutrino Majorana mass in the LRS Zee model~\cite{Khan:2018jge}.}
\label{fig:rhnmass}
\end{figure}
\end{center}

Fig.~\ref{fig:rhnmass} shows a plot of the Majorana masses of the right-handed neutrinos as generated at the one-loop level in this model. The plot shows the variation of the eigenvalues of the right-handed neutrino Majorana mass matrix (elements of the matrix given in Eq.~\ref{eq:RHM}) as a function of its coupling with the charged singlet scalar. As $\lambda'_R$ is varied from 0.1 to 3, the lightest right handed neutrino mass $M_{N_1}$ varies from 3 eV to 80 eV, while $M_{N_{2,3}}$ remain in the sub-MeV scale. In generating the plot, we considered the lightest charged Higgs boson has a mass 473 GeV, which primarily consists of charged singlet $\delta$. Below we discuss the contribution of the right-handed neutrinos in $0\nu \beta \beta$ decay.

\subsection{Diagrams and amplitudes of $0\nu \beta\beta$ transition}
\label{s:LRzee_neu02beta}

Contrary to most seesaw models which contain the right-handed neutrinos of TeV scale mass, the LRS Zee model naturally accommodates eV-MeV scale right-handed neutrinos, as has already been discussed in the previous section. Diagrams involving the right-handed neutrinos can thus significantly contribute to the $0\nu \beta\beta$ processes and this model gives us an excellent framework to study these effects.

\begin{figure}[h!]
\begin{center}
\subfloat[]{\includegraphics[scale=0.37]{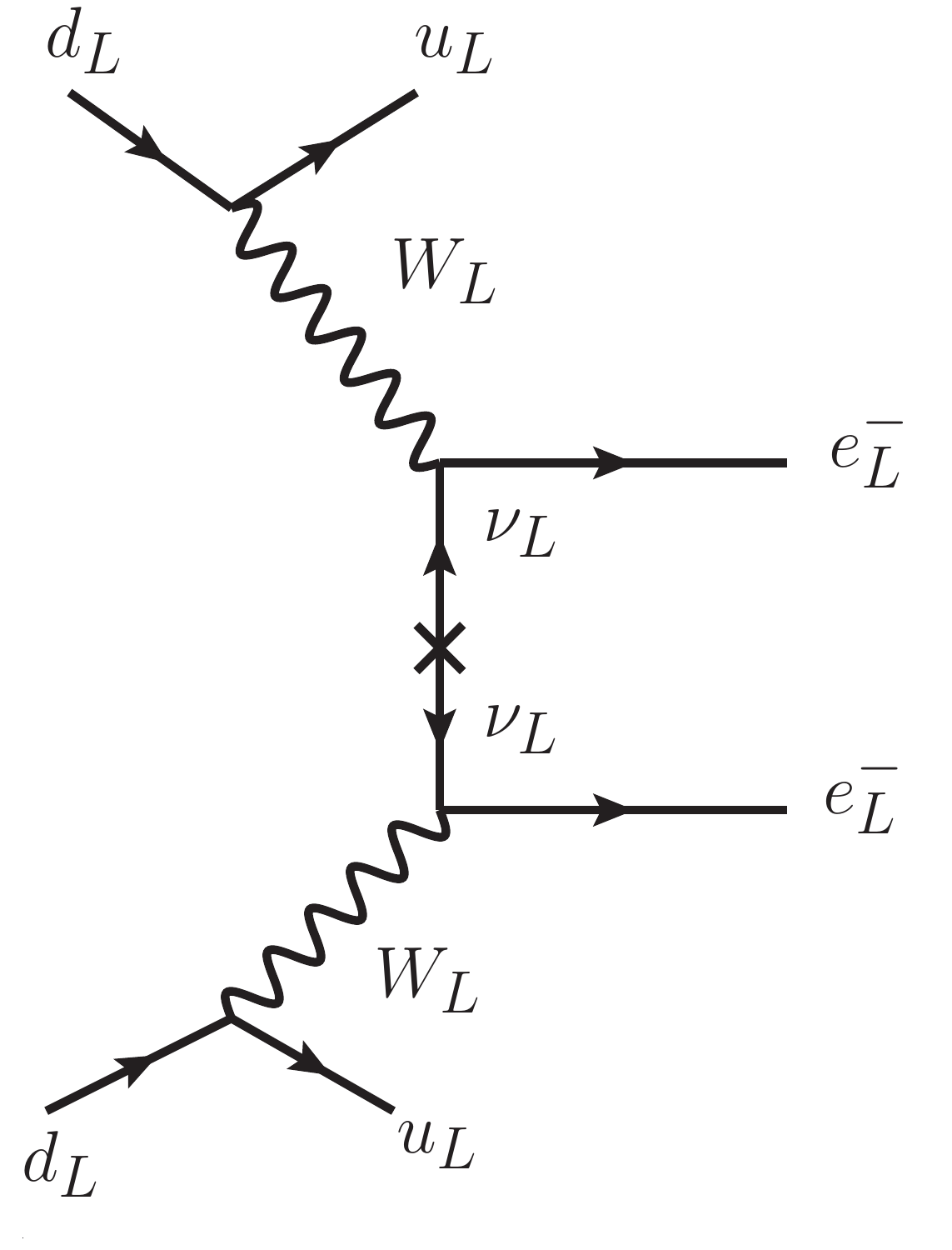}}
\subfloat[]{\includegraphics[scale=0.37]{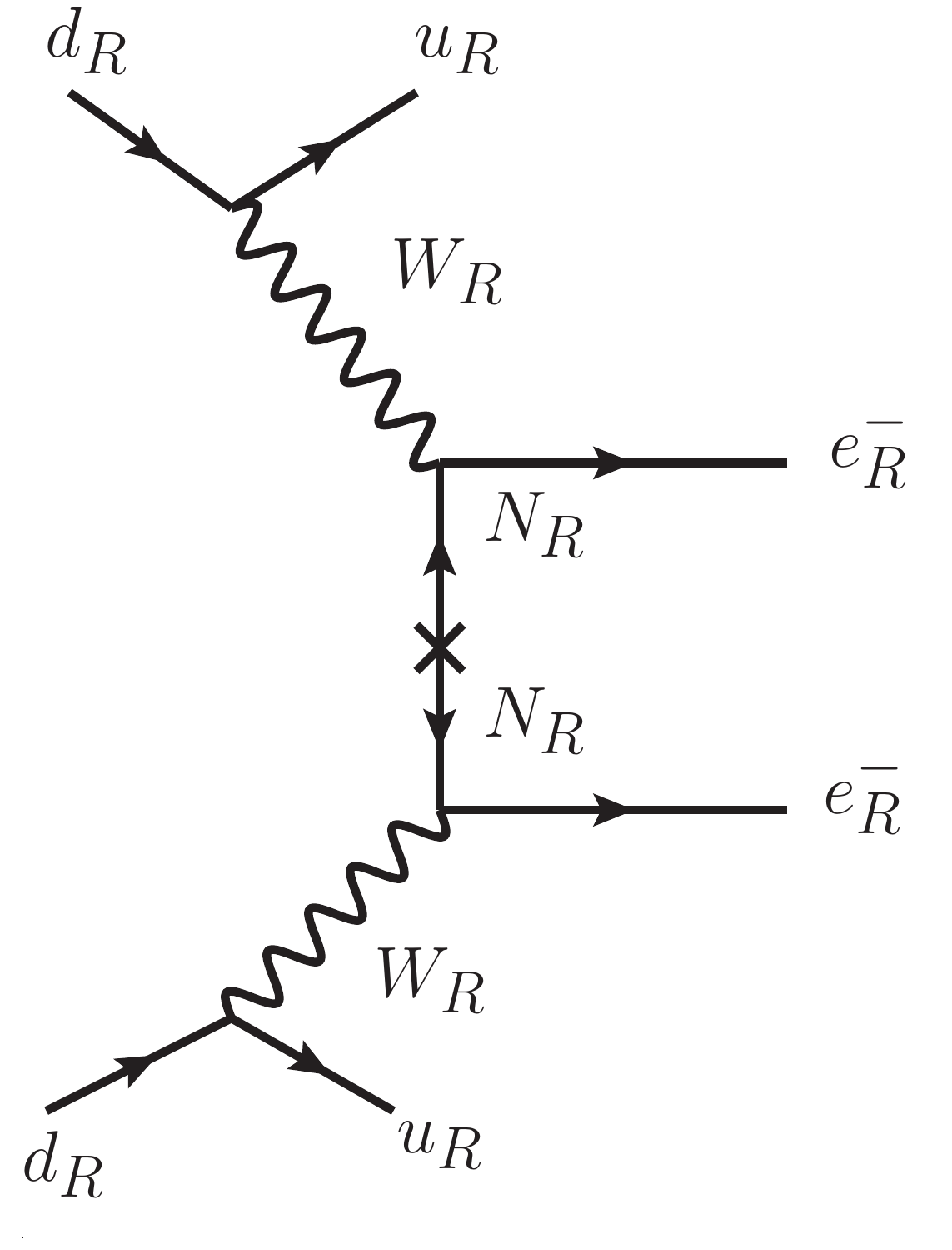}}
\subfloat[]{\includegraphics[scale=0.32]{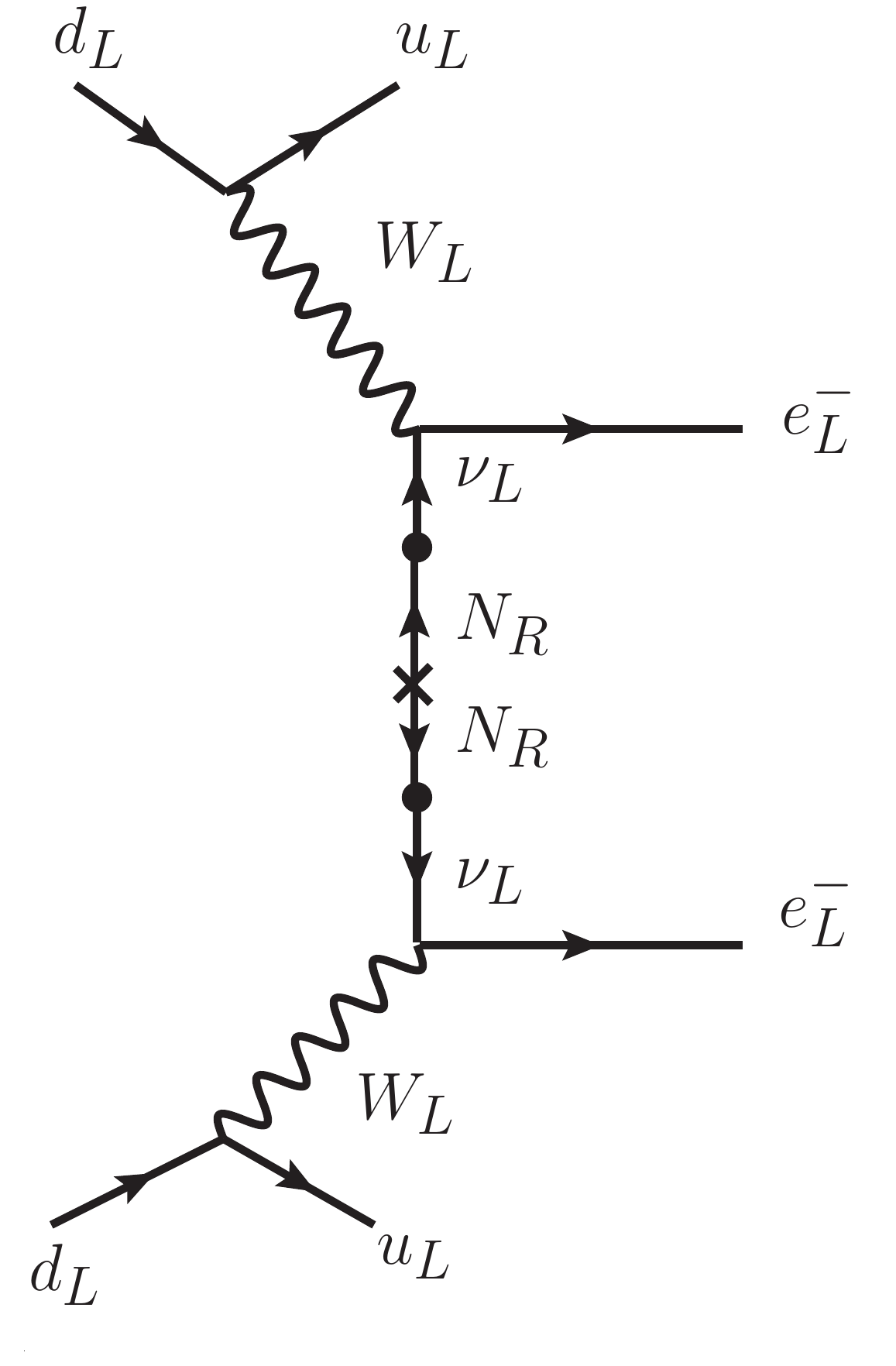}}
\subfloat[]{\includegraphics[scale=0.32]{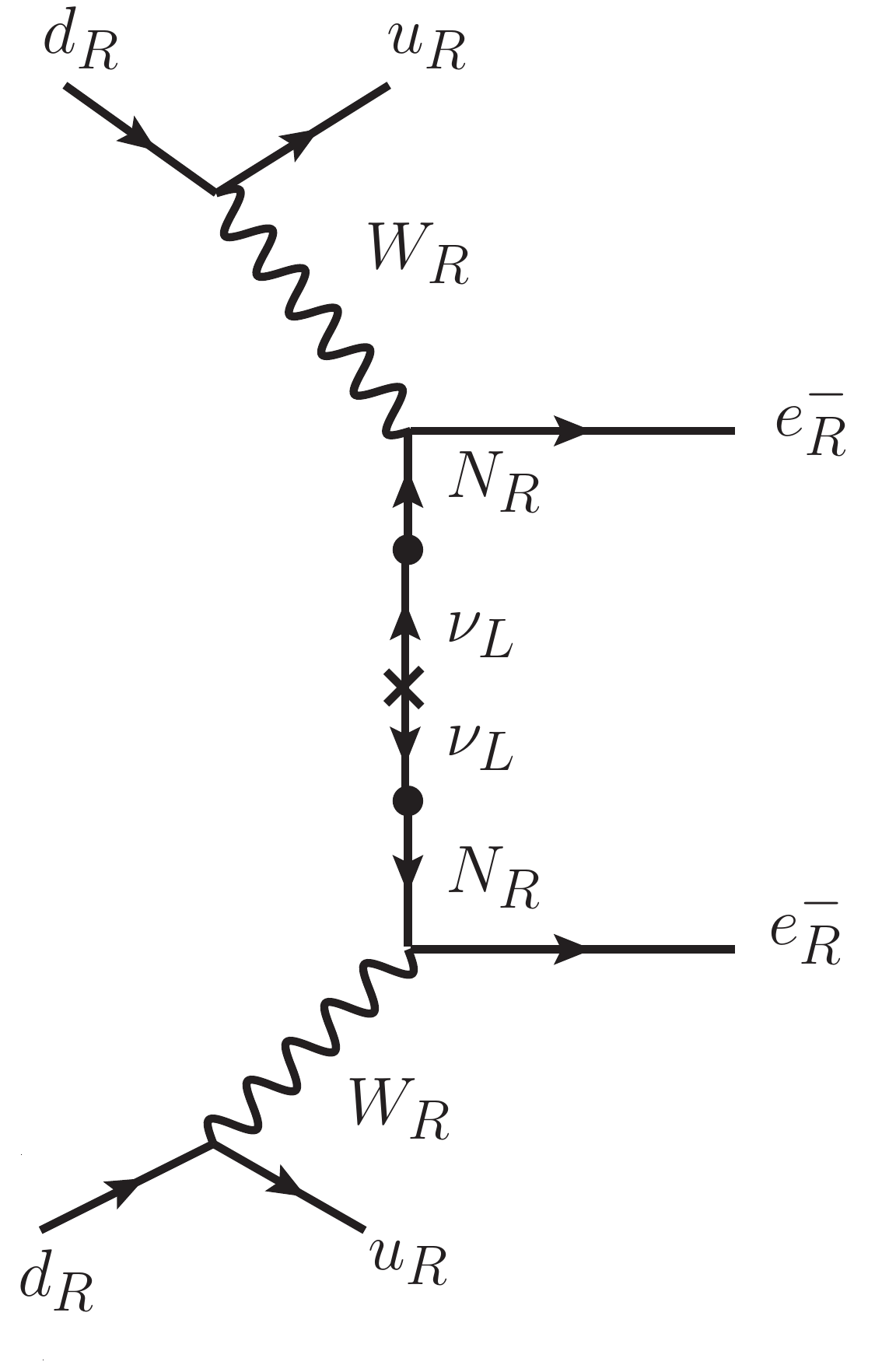}}
\end{center}
\begin{center}
\subfloat[]{\includegraphics[scale=0.37]{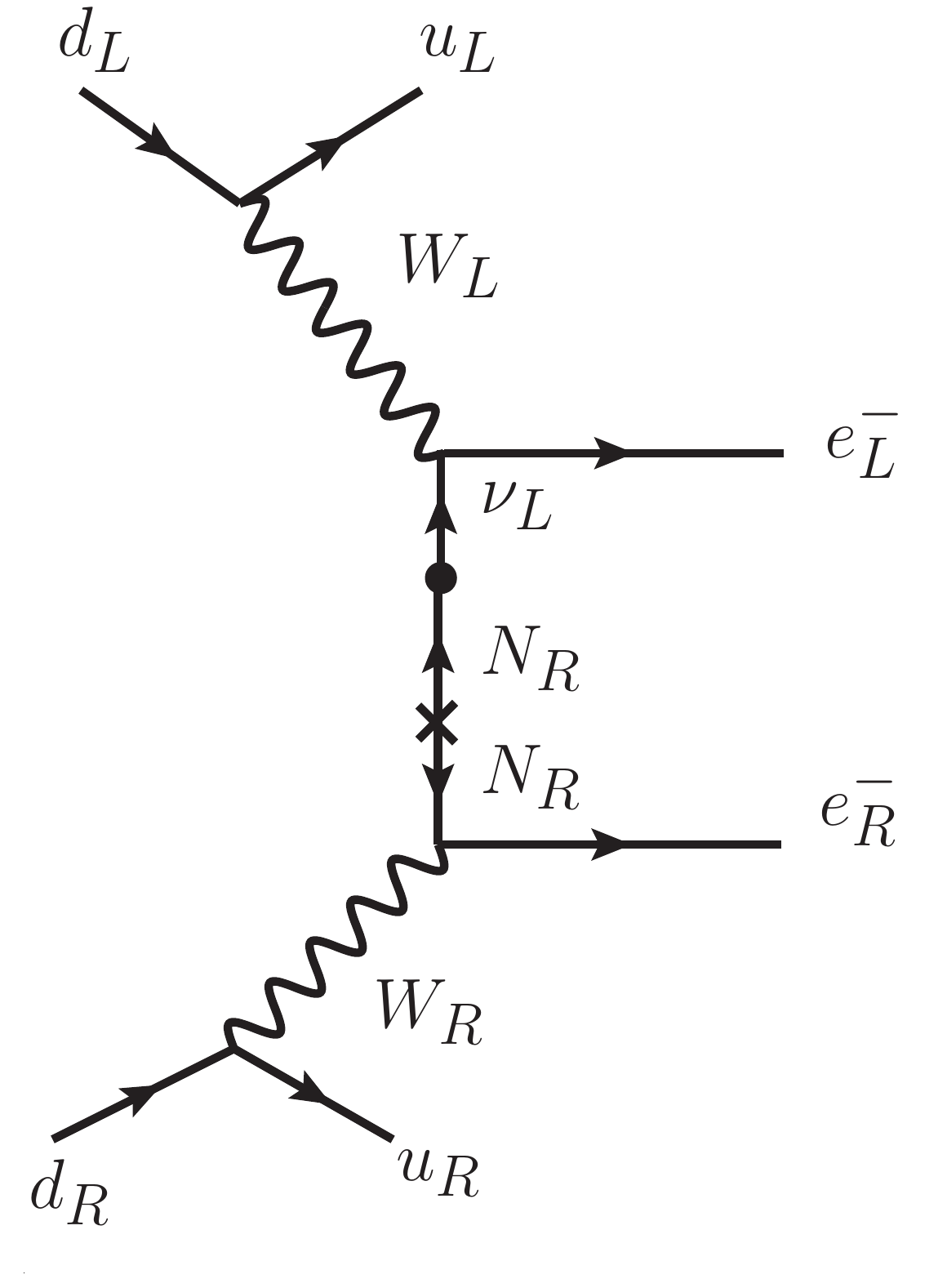}}
\subfloat[]{\includegraphics[scale=0.37]{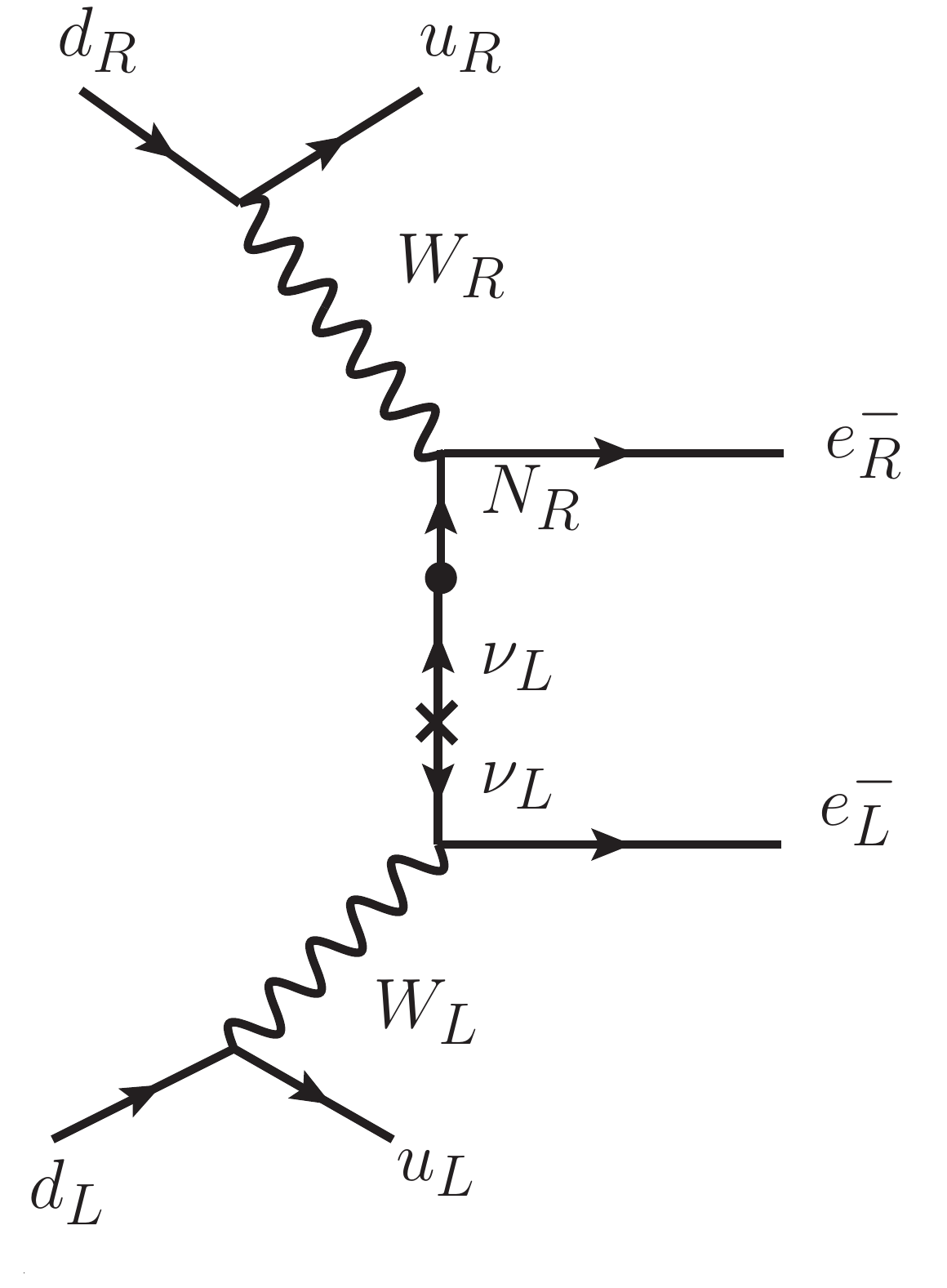}}
\subfloat[]{\includegraphics[scale=0.32]{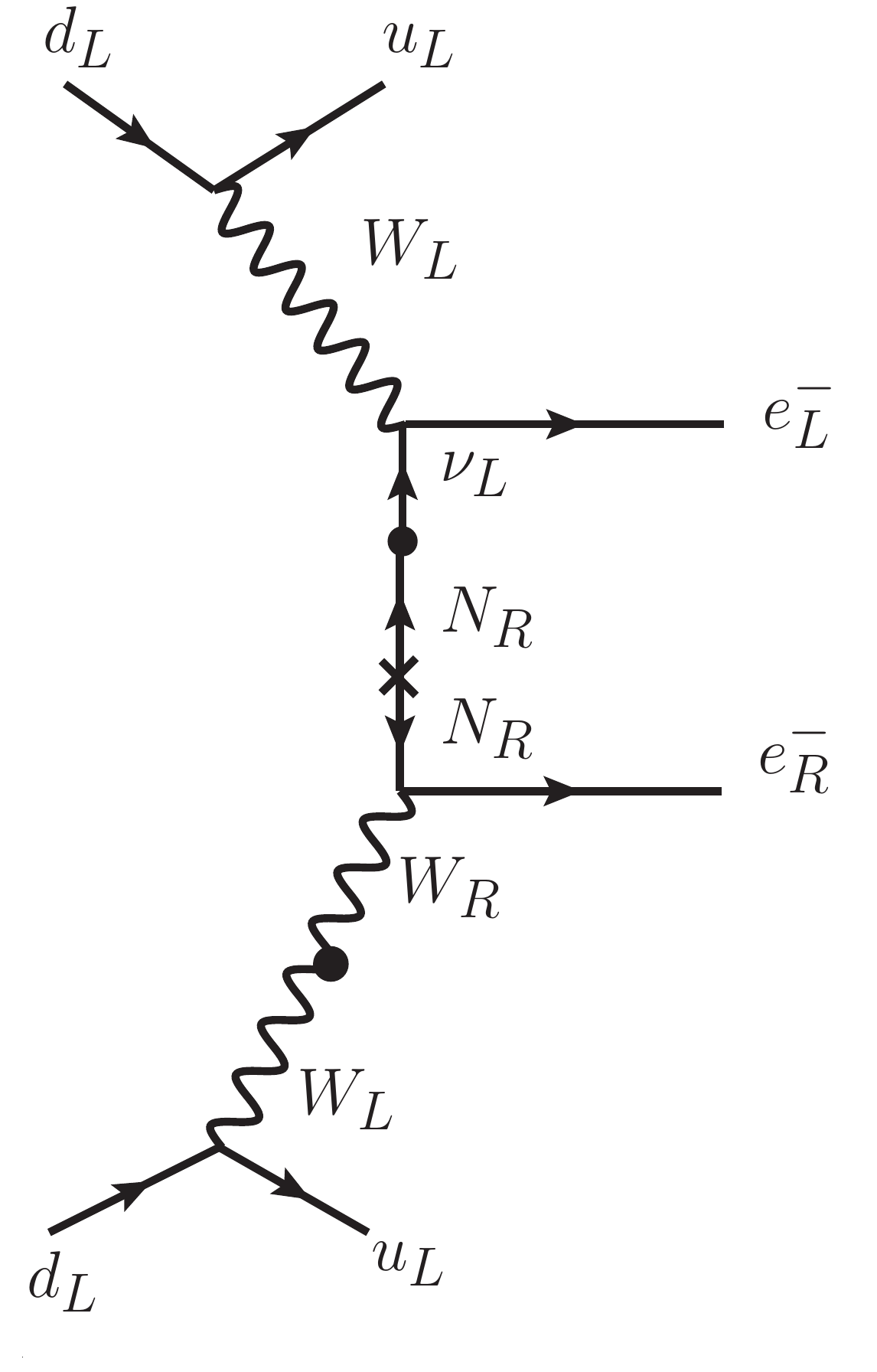}}
\subfloat[]{\includegraphics[scale=0.32]{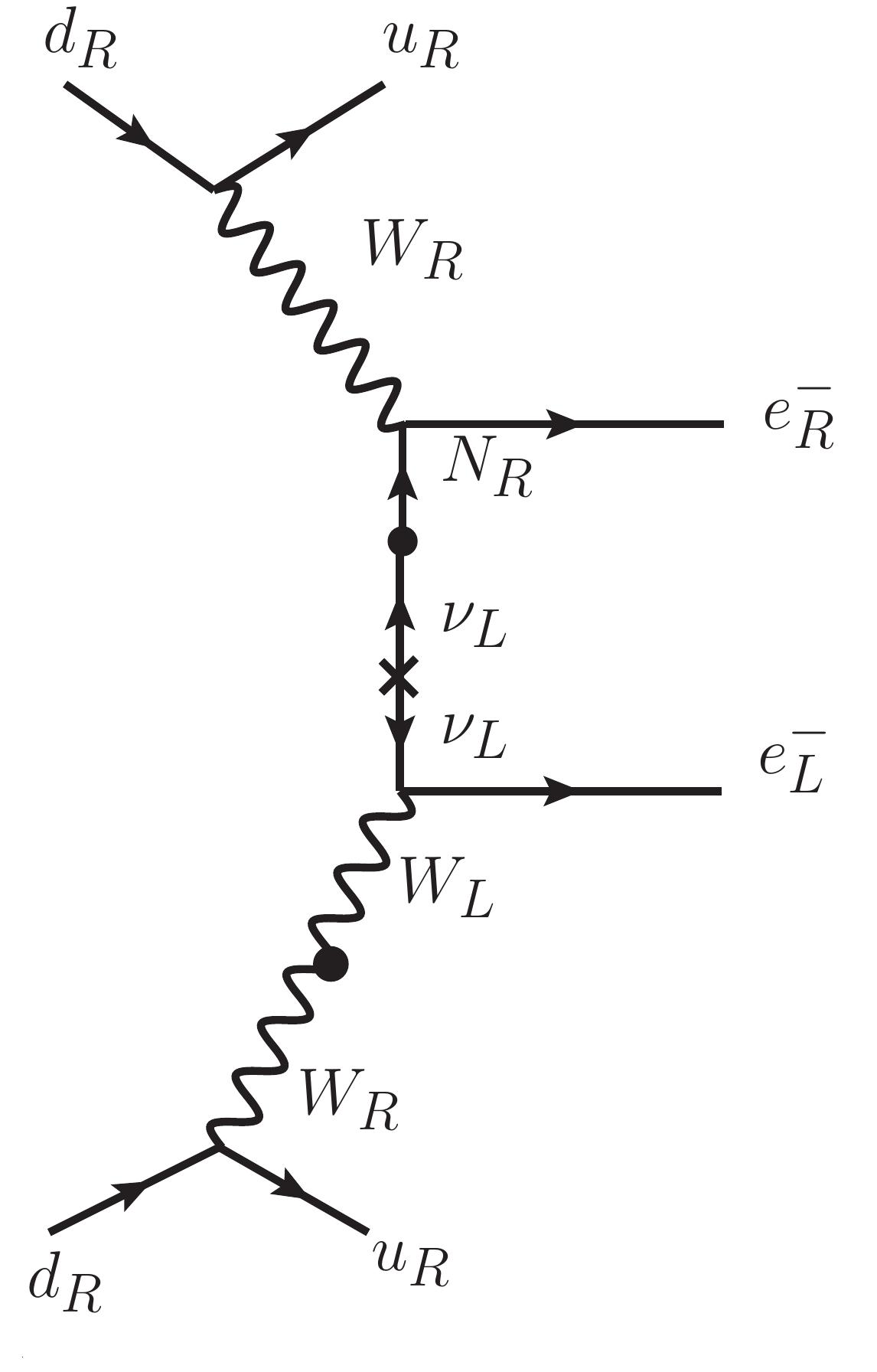}}
\caption{Feynman diagrams of all possible $0\nu \beta\beta$ processes in the LRS Zee model.}
\label{fig:feynman}
\end{center}
\end{figure}

The Feynman diagrams of all the possible contributions are presented in Fig~\ref{fig:feynman}. For each diagram we write its amplitude and identify the dimensionless parameter $\eta_i$ that will be used in the computation of the half life ($T_{1/2}^{0\nu}$) of  the $0\nu \beta\beta$ process. In the subsequent discussion, we refer the mass eigenstates of SM neutrinos as ‘light’, and the right-handed neutrinos as `heavy’, as the right-handed neutrino states are heavier than the SM neutrinos.

\begin{itemize}%[label={}]%(\alph*)]
\item{{\it Light neutrino diagram:} The diagram (a) corresponds to the light neutrino contribution. Its amplitude is given as
\begin{equation}
A_1 \simeq G_F^2 \sum_{i} \mathbb{U}_{ei}^2 \frac{m_i}{p^2}, 
\end{equation}
where $G_F$ is the Fermi constant, $p$ is the momentum transfer at the leptonic vertex and $i=1,2,3$ corresponds to the light neutrino mass eigenstates. The corresponding $\eta$ obtained in this case is given as
\begin{equation}
\eta_1 = \frac{1}{m_e}\sum_i \mathbb{U}_{ei}^2 m_i.
\end{equation}
}
\item{{\it Heavy neutrino diagrams:} Diagram (b) corresponds to the heavy neutrino contribution. The heavy neutrinos in this model are composed of the right-handed neutrinos but unlike other Left-Right models, they are quite light in this case with masses in the eV to MeV range. The Feynman amplitude and the corresponding $\eta$ from this diagram is given as
\begin{equation}
A_2 \simeq G_F^2 \( \frac{M_{W_L}}{M_{W_R}} \) ^4 \sum_{i} \mathbb{V}_{ei}^2 \frac{M_i}{p^2},~~~~
\eta_2 = \frac{1}{m_e} \( \frac{M_{W_L}}{M_{W_R}}\) ^4 \sum_i \mathbb{V}_{ei}^2 M_i,
\end{equation}
where the summation is over the  heavy neutrino eigenstates in this case. }
\item{{\it Light-heavy neutrino mixing diagram:} Diagrams (c)\,\&\,(d) correspond to the contributions due to the mixing between the light and heavy neutrinos. The Feynman amplitudes are given as
\begin{equation}
A_3 \simeq G_F^2 \sum_{i} \mathbb{S}_{ei}^2 \frac{M_i}{p^2},~~~~   A_4 \simeq G_F^2 \( \frac{M_{W_L}}{M_{W_R}} \) ^4 \sum_{i} \mathbb{T}_{ei}^2 \frac{m_i}{p^2} ,
\end{equation}
while the $\eta$ factors are
\begin{equation}
\eta_3 = \frac{1}{m_e}\sum_i \mathbb{S}_{ei}^2 M_i,~~~~\eta_4 = \frac{1}{m_e} \( \frac{M_{W_L}}{M_{W_R}}\) ^4 \sum_i \mathbb{T}_{ei}^2 m_i.
\end{equation}
} 
\item{{\it $\lambda$ diagrams:} Diagrams (e)\,\&\,(f) represent the processes mediated by the $W_L-W_R$ exchange. The Feynman amplitudes from each diagram can be easily combined to give us a final expression which is
\begin{equation}
A_{\lambda} \simeq G_F^2 \( \frac{M_{W_L}}{M_{W_R}} \)^2 \sum_i \frac{ \mathbb{U}_{ei} \mathbb{T}_{ei}^* + \mathbb{V}_{ei} \mathbb{S}_{ei}^* }{p}, 
\end{equation}
and the expression for the $\eta$ parameter is
\begin{equation}
\eta_{\lambda} = \( \frac{M_{W_L}}{M_{W_R}} \)^2 \sum_i  \mathbb{U}_{ei} \mathbb{T}_{ei}^* + \mathbb{V}_{ei} \mathbb{S}_{ei}^* .
\end{equation}
 }
 \item{{\it $\eta$ diagrams:} Diagrams (g)\,\&\,(h) are due to the $W_L-W_R$ mixing in this model and depend on the $W_L-W_R$ mixing angle $\theta_{LR}$. The Feynman amplitude combining the two diagrams can be written as
\begin{equation}
A_{\eta} = G_F^2 \tan \theta_{LR} \sum_i \frac{\mathbb{U}_{ei} \mathbb{T}_{ei}^*+ \mathbb{V}_{ei} \mathbb{S}_{ei}^*}{p}
\end{equation} 
and the corresponding $\eta$ parameter is 
\begin{equation}
\eta_{\eta} = \tan \theta_{LR} \sum_i \mathbb{U}_{ei} \mathbb{T}_{ei}^*+ \mathbb{V}_{ei} \mathbb{S}_{ei}^*.
\end{equation}
  }
\end{itemize}
The half-life for the $0\nu \beta\beta$ process after combining the contributions from all these diagrams is then given as~\cite{Barry:2013xxa,Awasthi:2013ff,Chakrabortty:2012mh,BhupalDev:2014qbx}
\begin{equation}
{T_{1/2}^{0\nu}} = \[ G_{0\nu} \( \left| M_{\nu}^{0\nu} \eta_1 + M_{\nu}^{0\nu} \eta_3 \right|^2 +\left| M_{\nu}^{0\nu} \eta_2+M_{\nu}^{0\nu} \eta_4 \right|^2 + \left| M_{\lambda}^{0\nu} \eta_{\lambda} + M_{\eta}^{0\nu} \eta_{\eta} \right|^2\) \]^{-1},
\label{eq:thalf}
\end{equation}
where $G_{0\nu}$ is the phase space factor; $M_{\nu}^{0\nu}$, $M_{\lambda}^{0\nu}$ and $M_{\eta}^{0\nu}$ are the nuclear matrix elements.

Now that we have the expression for the half-life for the $0\nu \beta\beta$ processes, let us discuss some of the features of this framework which will help us understand the relative contribution arising from each of these diagrams. The right-handed neutrino masses being at the eV to MeV scale contribute significantly to these processes here and hence the diagrams involving $N_R$ become quite important. The relative contributions from the diagrams are also highly dependent on the light-heavy neutrino mixings ($\mathbb{S}$, $\mathbb{T}$) with the $\lambda$ and $\eta$ diagrams becoming significant as this mixing increases. The gauge boson ($W_L - W_R$) mixing is another important factor in these diagrams and its value can determine which diagram gives significant contribution to the $0\nu \beta\beta$ decay process. Finally since the $W_R$ boson mass is required to be quite large from experimental constraints \cite{Mohapatra:1983ae,Pospelov:1996fq,Zhang:2007da,Maiezza:2010ic,Chakrabortty:2012pp,Aad:2015xaa,Kaya:2015tia,Arkani-Hamed:2015vfh,Golling:2016gvc,Mitra:2016kov,Borah:2017leo,Cai:2017mow,Beringer:1900zz}, we have chosen it to be $5.5$\,TeV here and this results in a large suppression for all the diagrams with amplitudes involving $ \( M_{W_L}/{M_{W_R}} \right)^4$ term. 

The neutrino parameters in this model depend significantly on the masses and mixings of the charged scalars as can be seen quite clearly from Eq.~\ref{eq:RHM}. A close inspection of this equation also shows that the factor $V_{5i}$ which is the mixing between the charged singlet and other charged Higgs states is quite important for the neutrino masses. As discussed earlier, the charged singlet Higgs $\delta^+$ can only have significant mixing with the left-handed charged scalar $H_L^+$. We can thus approximately write 
\begin{eqnarray}
H_1^+ &=& \delta^+ \cos \theta  + H_L^+ \sin \theta , \notag \\
H_2^+ &=& -\delta^+ \sin \theta  + H_L^+  \cos \theta,
\end{eqnarray} 
where $H_1^+$ and $H_2^+$ are the lightest and next-to-lightest charged Higgs bosons respectively with $\theta$ being the mixing angle. Clearly two extreme cases appear here
\begin{enumerate}
\item{maximal mixing with $\theta$ = $\pi/4$, denoted as ${\text H_{\text {max}}}$},
\item{minimal mixing with $\theta$ = $0$, denoted as ${\text H_{\text {min}}}$}.
\end{enumerate}
 
\subsection{Results}
\label{s:LRZee_result}

To analyse the half-life of $0\nu \beta \beta$ process for Germanium ($^{76}$Ge) and Xenon ($^{136}$Xe) nucleus, we consider two cases of maximal and minimal mixing as described in the previous section. We consider a normal mass ordering among  the SM neutrinos, and use the Casas-Ibarra~\cite{Casas:2001sr} parametrization to fit the  latest neutrino oscillation data~\cite{Esteban:2020cvm}.  As the right-handed neutrino masses are quite small here, the mixing between them and the left-handed neutrinos, represented by the $\mathbb{S}$ and $\mathbb{T}$ matrices in Eq.~\ref{eq:numix}, can become quite significant\footnote{These terms still remain orders of magnitude smaller than the $\mathbb{U}$ and $\mathbb{V}$ matrices.}. For a fixed choice of the right-handed neutrino masses, the light-heavy neutrino ($\nu_L - N_R$) mixing depends largely on the light neutrino masses. As the SM neutrino masses  increase, the mass difference between the light and heavy neutrino  states become smaller resulting in a larger mixing angle. Thus the lightest neutrino mass $m_{\nu_1}$ is an important parameter for our analysis. 

The other parameters which play a significant role in determining the value of $T_{1/2}^{0\nu}$ are the $W_L-W_R$ mixing angle ($\theta_{LR}$) and the Dirac CP phase ($\delta_{CP}$) of the neutrino Pontecorvo-Maki-Nakagawa-Sakata (PMNS) matrix. The contribution from the $\eta$ diagram, directly proportional to $\tan \theta_{LR}$, can become substantial depending on this mixing angle. The value of $\delta_{CP}$, though directly does not appear in any of the expressions, determines the neutrino parameters obtained from the Cassas-Ibarra parametrization. This has a significant consequence on the calculated value of $T_{1/2}^{0\nu}$. The nuclear matrix elements (NMEs) for $^{76}$Ge and $^{136}$Xe, which we adopt from \cite{Agostini:2018tnm,KamLAND-Zen:2016pfg} are equally important for the evaluation of $T^{0\nu}_{1/2}$. We consider two cases, one with the maximum and another with the minimum values of the NMEs, and evaluate the half-life. For each nucleus ($^{76}$Ge or $^{136}$Xe), we thus get four separate cases which are:
\begin{enumerate}[(a)]
\item {\bf ${{\text H_{\text {min}}}-{\text {($^{76}$Ge/$^{136}$Xe)}}_{\text {min}}}$:} Corresponds to the case where the Higgs boson mixing is minimum and minimum value for the ${}^{76}$Ge/${}^{136}$Xe NME has been used.
\item {\bf ${{\text H_{\text {max}}}-{\text {($^{76}$Ge/$^{136}$Xe)}}_{\text {min}}}$}: Corresponds to the case where the Higgs boson mixing is maximum and minimum value for the ${}^{76}$Ge/${}^{136}$Xe NME has been used.
\item {\bf ${{\text H_{\text {min}}}-{\text {($^{76}$Ge/$^{136}$Xe)}}_{\text {max}}}$:} Corresponds to the case where the Higgs boson mixing is minimum and maximum value for the ${}^{76}$Ge/${}^{136}$Xe NME has been used.
\item {\bf ${{\text H_{\text {max}}}-{\text {($^{76}$Ge/$^{136}$Xe)}}_{\text {max}}}$:} Corresponds to the case where the Higgs boson mixing is maximum and maximum value for the ${}^{76}$Ge/${}^{136}$Xe NME has been used. 
\end{enumerate} 
 For each of these cases we vary $m_{\nu_1}$, $\theta_{LR}$ and $\delta_{CP}$ to obtain the predicted value of the half-life of  $0\nu \beta\beta$ decay process. 
\begin{figure}[h!]
\begin{center}
\includegraphics[scale=0.45]{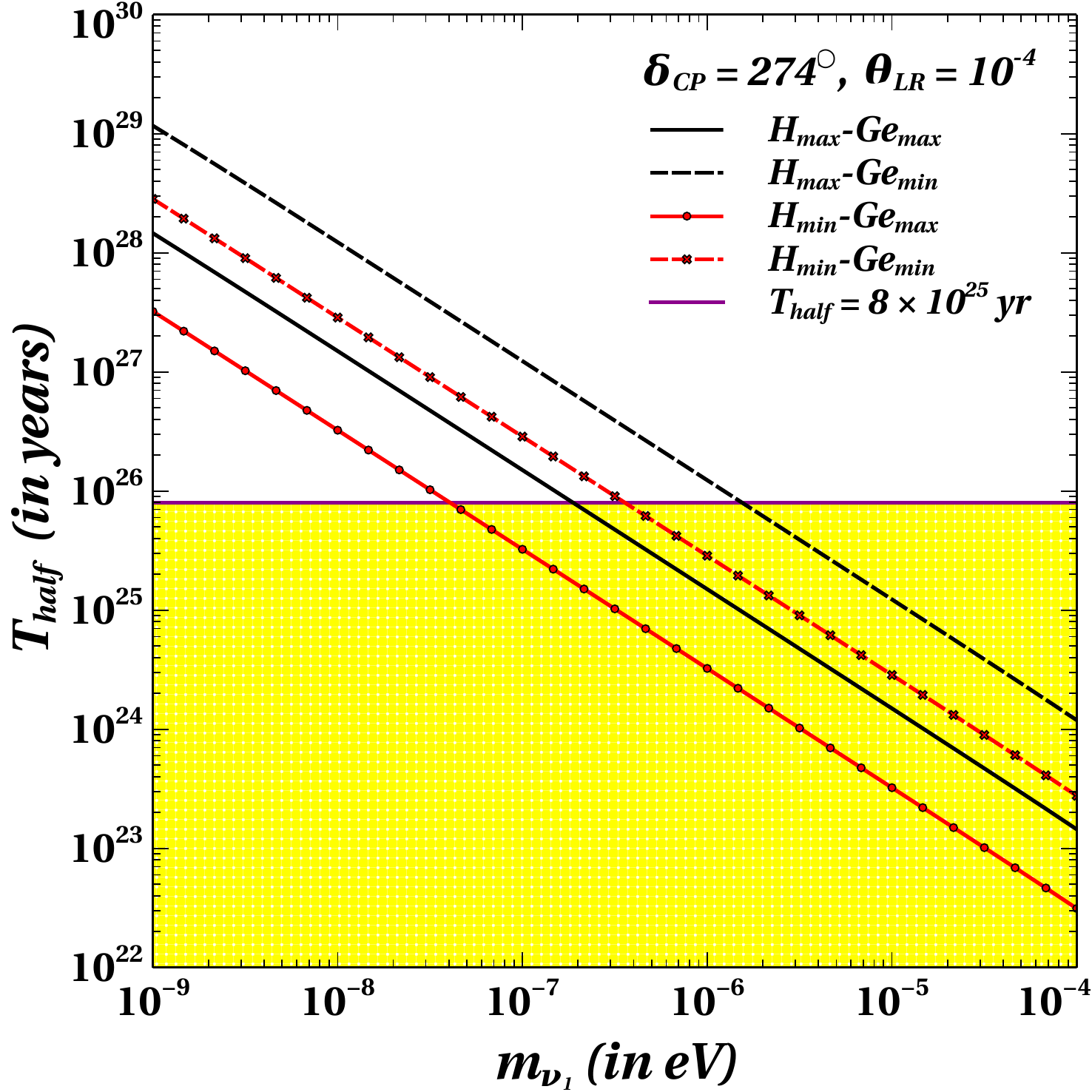}~~~~~~~~~~
\includegraphics[scale=0.45]{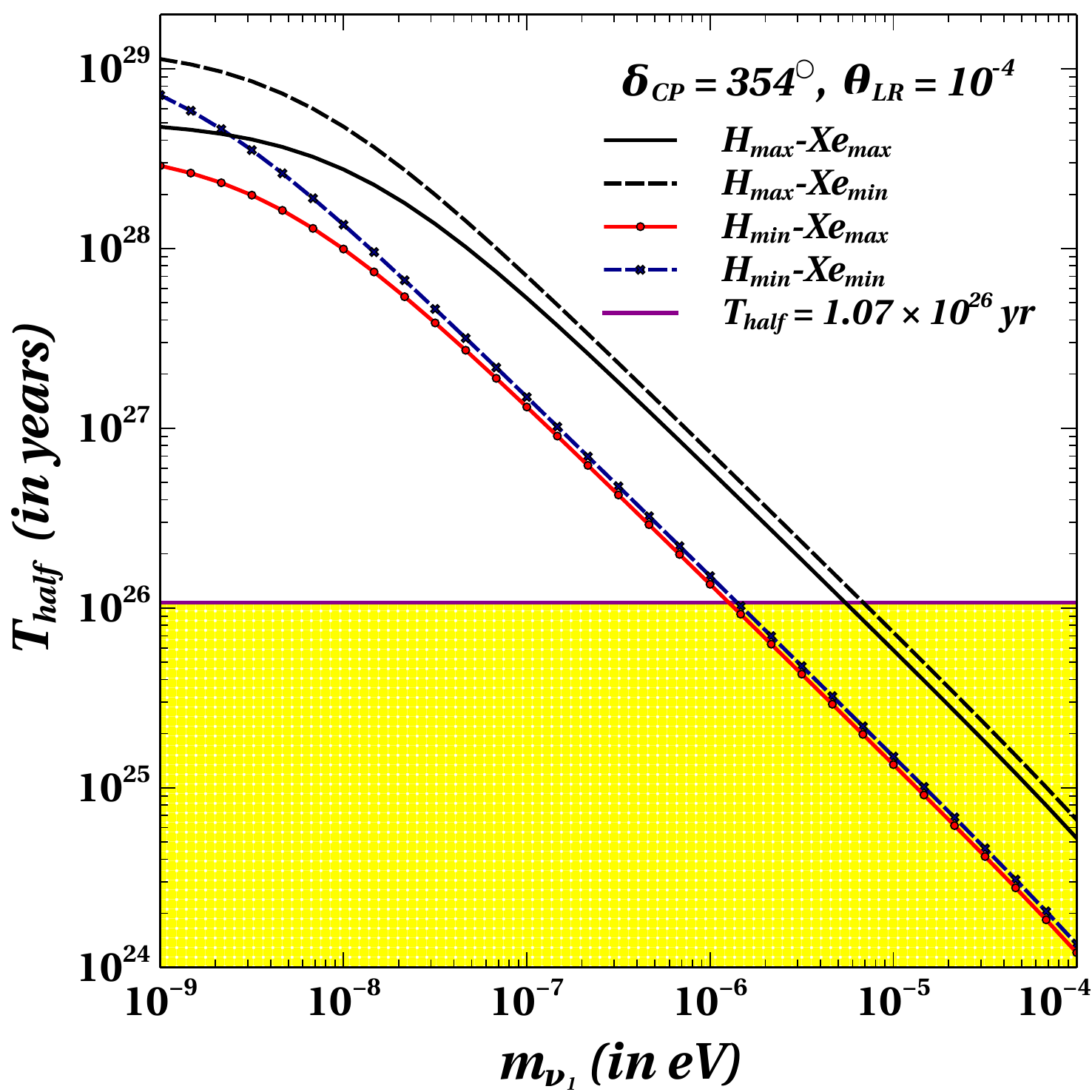}
\caption{Half-life of $0\nu \beta\beta$ process for $^{76}$Ge  and $^{136}$Xe nucleus as a function of lightest neutrino mass. The shaded region correspond to $T^{0\nu}_{1/2}< 8.0\times 10^{25}$ years for left panel, and $1.07\times 10^{26}$ years for right panel and disallowed by GERDA \cite{Agostini:2018tnm}, and KamLAND-Zen \cite{KamLAND-Zen:2016pfg}, respectively. }
\label{fig:tgemnu}
\end{center} 
\end{figure}
%Considering all these factors, we calculate the total decay width for each nucleus (Xe and Ge) for all the four cases discussed above. 
Fig.~\ref{fig:tgemnu} shows the variation of $T^{0\nu}_{1/2}$  for $^{76}$Ge nucleus as a function of the lightest neutrino mass $m_{\nu_1}$ for a fixed value of $\theta_{LR}$ and $\delta_{CP}$. The values of all other PMNS matrix elements were fixed to their central values and the $\lambda'_R$ matrix was chosen such that the right-handed neutrino masses were 7.92 eV, 3.54 keV and 3.55 keV respectively. As can be seen here, the half-life decreases quite drastically as the lightest neutrino mass increases. This is because the light-heavy neutrino mixing increases as discussed earlier and as a result the $\eta_3,~\eta_{\lambda}$ and $\eta_{\eta}$ contributions become dominant. As for this figure, we consider a large value of $\theta_{LR}$, therefore $\eta_{\eta}$ always dominate. We find that the canonical light neutrino contribution $\eta_1$ is rather subdominant in this figure.

\begin{figure}[h!]
 \begin{center}
\includegraphics[scale=0.45]{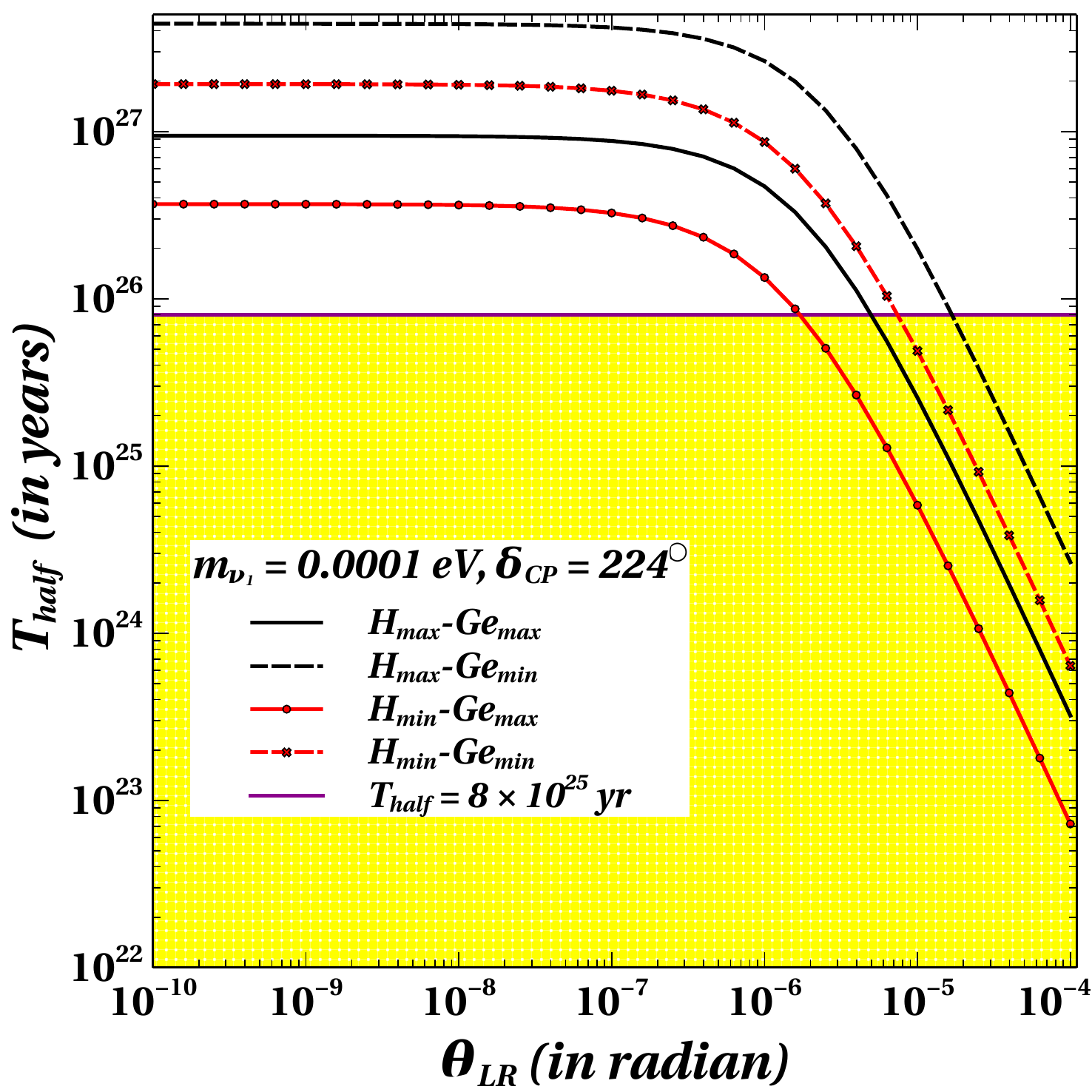}~~~~~~~~~~
\includegraphics[scale=0.45]{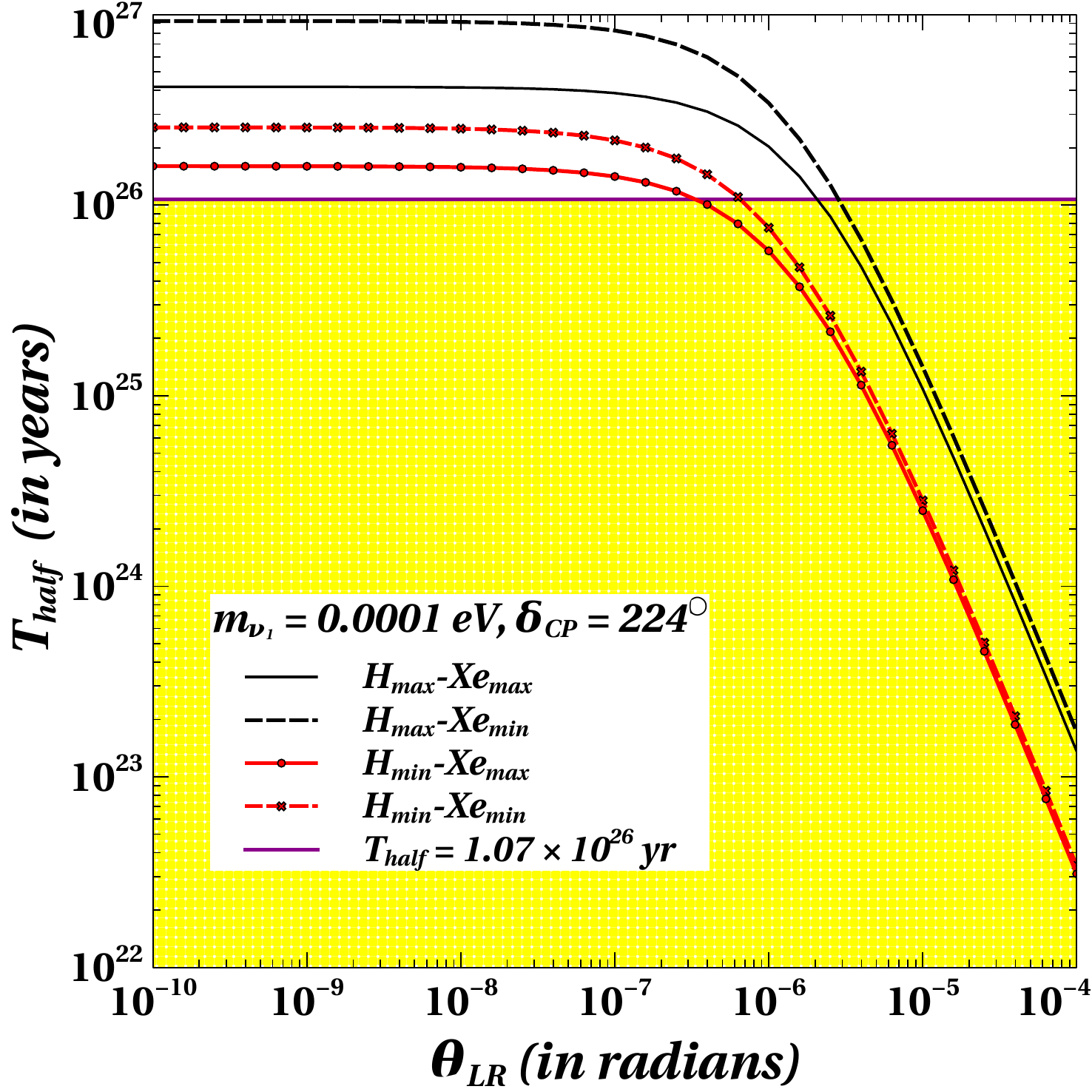}
\caption{Half-life of $0\nu \beta\beta$ process for $^{76}$Ge (left panel) and $^{136}$Xe nucleus (right panel) as a function of left-right charged gauge boson mixing.}
\label{fig:tgetheta}
\end{center}
\end{figure}

\begin{figure}[h!]
\begin{center}
\includegraphics[scale=0.45]{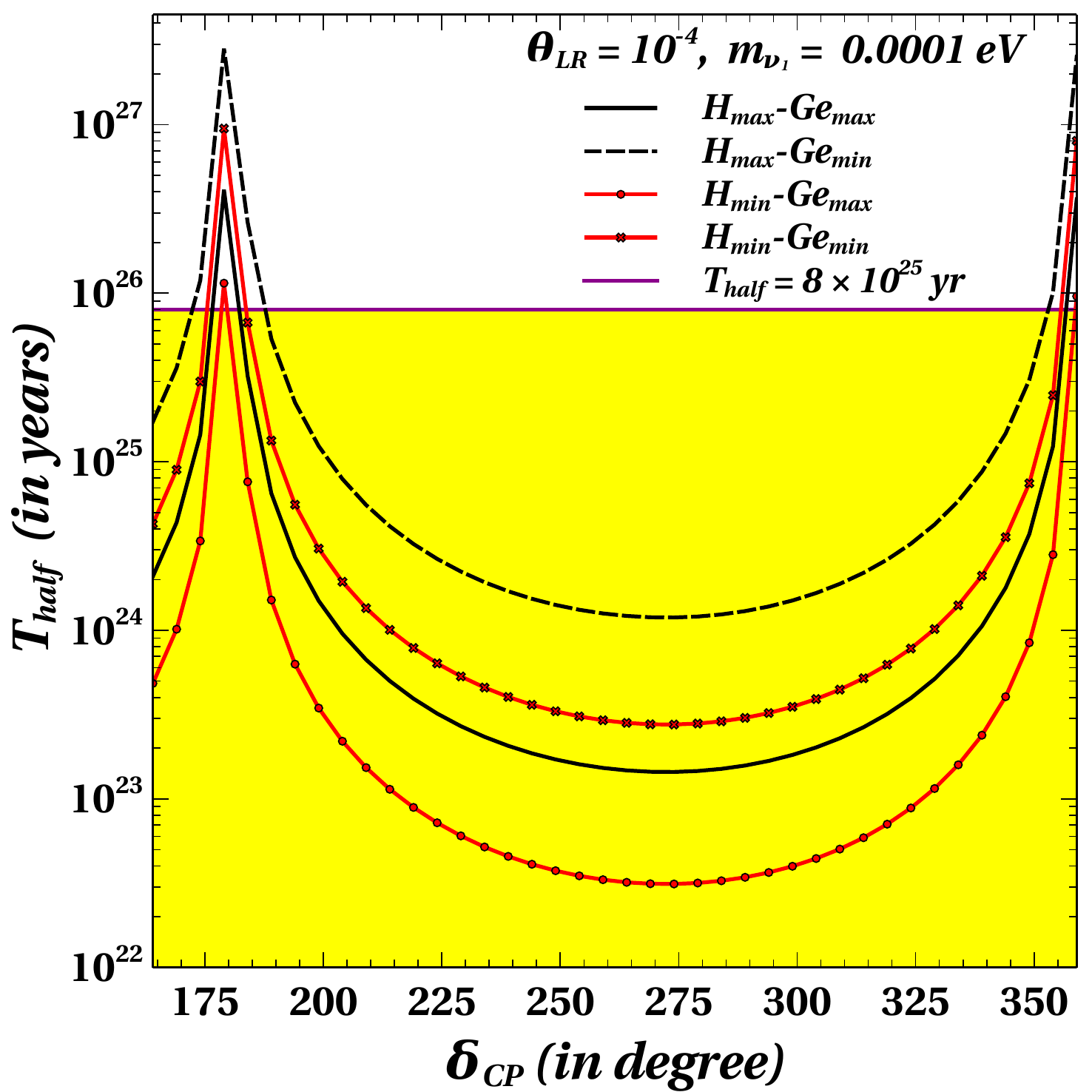}~~~~~~~~~~
\includegraphics[scale=0.45]{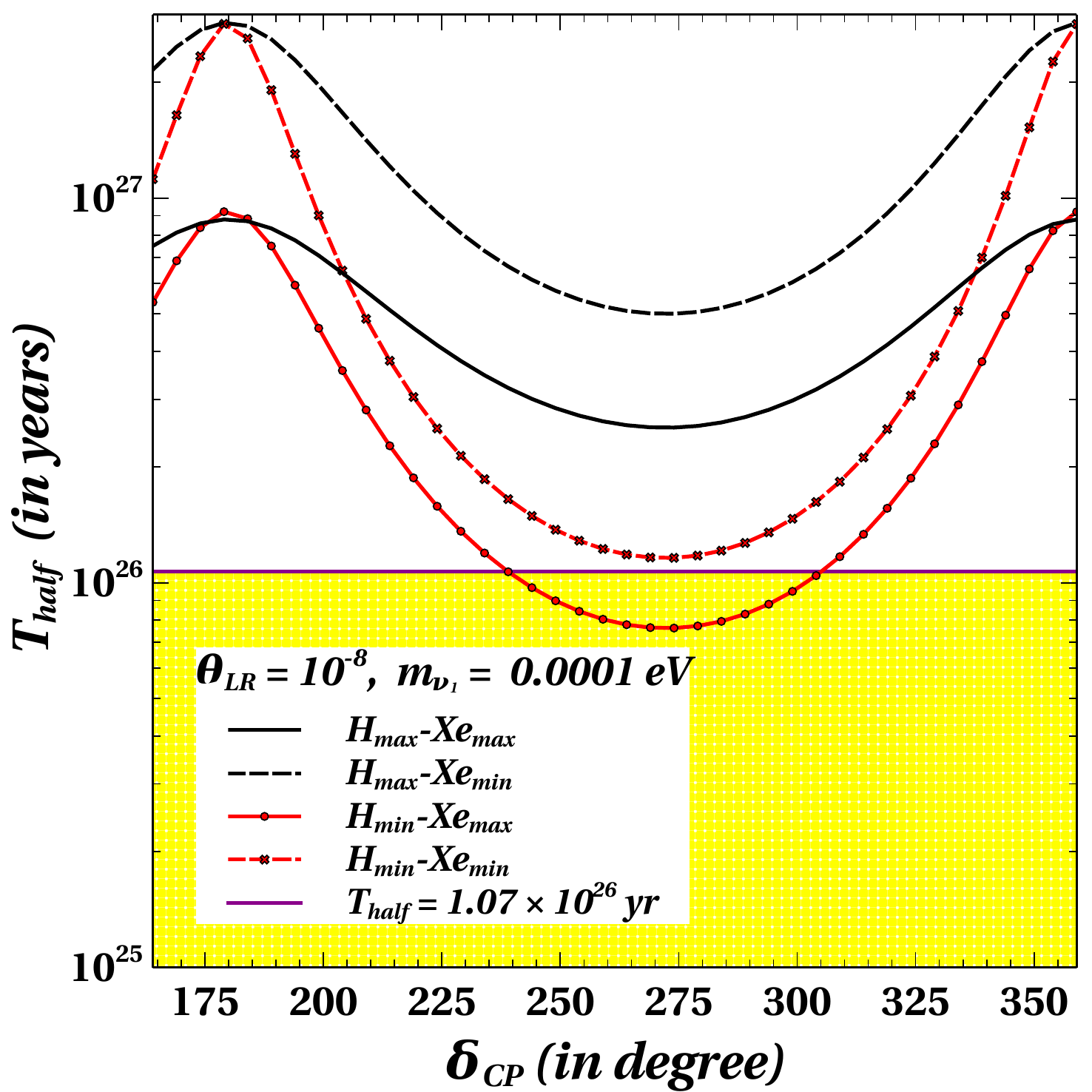}
\caption{Half-life of $0\nu \beta\beta$ process for ${}^{76}$Ge and ${}^{136}$Xe nucleus as a function of CP violating phase in the PMNS matrix.}
\label{fig:tgedcp}
\end{center}
\end{figure}

 Fig.~\ref{fig:tgetheta} shows the variation of $T^{0\nu}_{1/2}$ with the left-right charged gauge boson mixing $\theta_{LR}$. As $\theta_{LR}$ increases the decay half-life falls drastically for a value  $\theta_{LR} \gtrsim 10^{-6}$. This is the point at which the $\eta_{\eta}$ term, which is  proportional to $\tan \theta_{LR}$ starts dominating over the other terms resulting in a steep decrease of the half-life as expected. For smaller value of $\theta_{LR}$, the dominant contribution arises from $\eta_1,\eta_3,\eta_{\lambda}$, which are independent of the left-right mixing. Since the CP violating phase $\delta_{CP}$ is another crucial parameter in our analysis, we show the variation of half-life with respect to $\delta_{CP}$. Fig.~\ref{fig:tgedcp} gives the change of $T_{1/2}^{0\nu}$ for $^{76}$Ge nucleus as a function of $\delta_{CP}$.  A close inspection of the numbers we obtained shows that the variation of half-life mirrors the variation in $\sin \delta_{CP}$ which directly determines the values of the neutrino parameters obtained in our calculations. This is to note  that,  in all these figures, the scenario  ${{\text H_{\text {min}}}-{\text {Ge}}_{\text {max}}}$ gives the  strongest constraint. This can be understood from  the expression of  $T^{0\nu}_{1/2}$ as given in Eq.~\ref{eq:thalf}. As can be seen,  the calculated half-life is  smaller (leading to a more constrained scenario) for larger values of the amplitudes and NMEs. Therefore, naturally the  maximum values of $^{76}$Ge NMEs leads to a more constrained scenario. 
 
 The charged Higgs mixing on the other hand plays an indirect but significant role in determining the values obtained for the Feynman diagram amplitudes. As was discussed earlier, the Feynman amplitudes corresponding to $\eta_2$,  and $\eta_4$ are negligible due to the $ \( M_{W_L}/{M_{W_R}} \right)^4$ suppression. So the dominant contribution always arises from any one of the $\eta_1$, $\eta_\lambda$ or $\eta_\eta$ terms, while we find $\eta_3$ contribution is slightly smaller than these above mentioned contributions. A smaller charged Higgs mixing will invariably lead to a lighter Majorana mass for the RH neutrinos, which has a two-fold effect on the neutrino sector. Firstly, lighter RH neutrinos will results in relatively heavier active neutrinos since the active neutrino mass is obtained by seesaw mechanism in our case. This will boost the $\eta_1$ amplitude resulting in a smaller value of $T_{1/2}$. Secondly, a heavier active neutrino will result in a larger light-heavy mixing as discussed earlier. This again helps boost the $\eta_\lambda$ and $\eta_\eta$ amplitudes further lowering the calculated half-life of $0\nu \beta \beta$ process, leading to a tight constraint on the parameter.

%The plots obtained for the Xe nucleus are very similar in nature to the ones for Ge nucleus and hence have not been included here. 
The plots obtained for the $^{136}$Xe nucleus are very similar in nature to the ones for $^{76}$Ge and the most constrained scenario is again the ${{\text H_{\text {min}}}-{\text {Xe}}_{\text {max}}}$ case. This warrants for a more detailed study of this particular case for a better understanding, which we discuss below. We present the results for both $^{76}$Ge and $^{136}$Xe nucleus in the ensuing discussion of the most constrained scenario for each, i.e., largest values for the NMEs and minimal mixing of the Higgs sector ${{\text H_{\text {min}}}-{\text {Ge}}_{\text {max}}}$ and ${{\text H_{\text {min}}}-{\text {Xe}}_{\text {max}}}$.

\begin{figure}[h!]
\begin{center}
\includegraphics[scale=0.45]{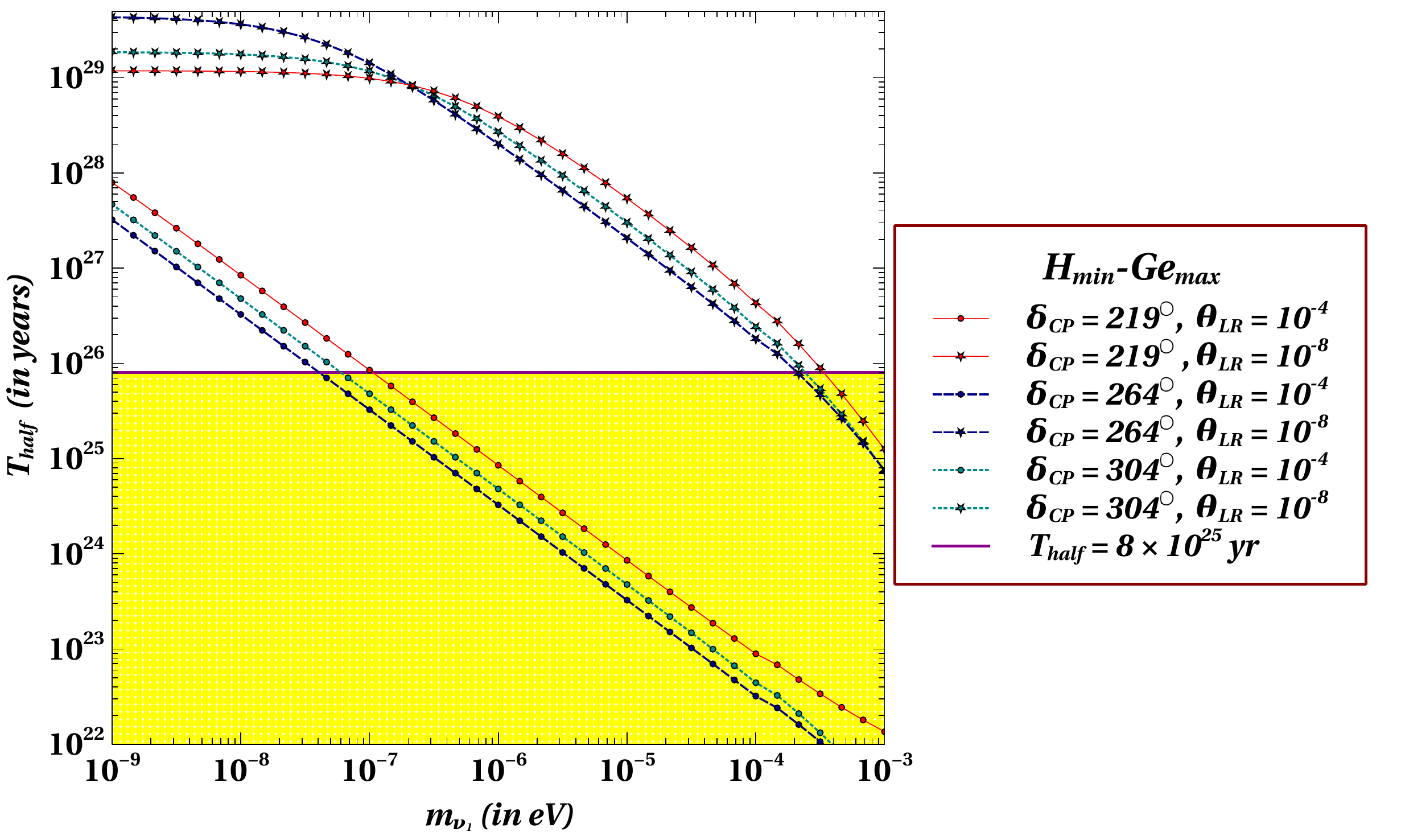}
\caption{Half-life of $0\nu \beta\beta$ process for ${}^{76}$Ge nucleus as a function of the the lightest neutrino mass for several fixed values of $\theta_{LR}$ and $\delta_{CP}$.}
\label{fig:gemnu}
\includegraphics[scale=0.45]{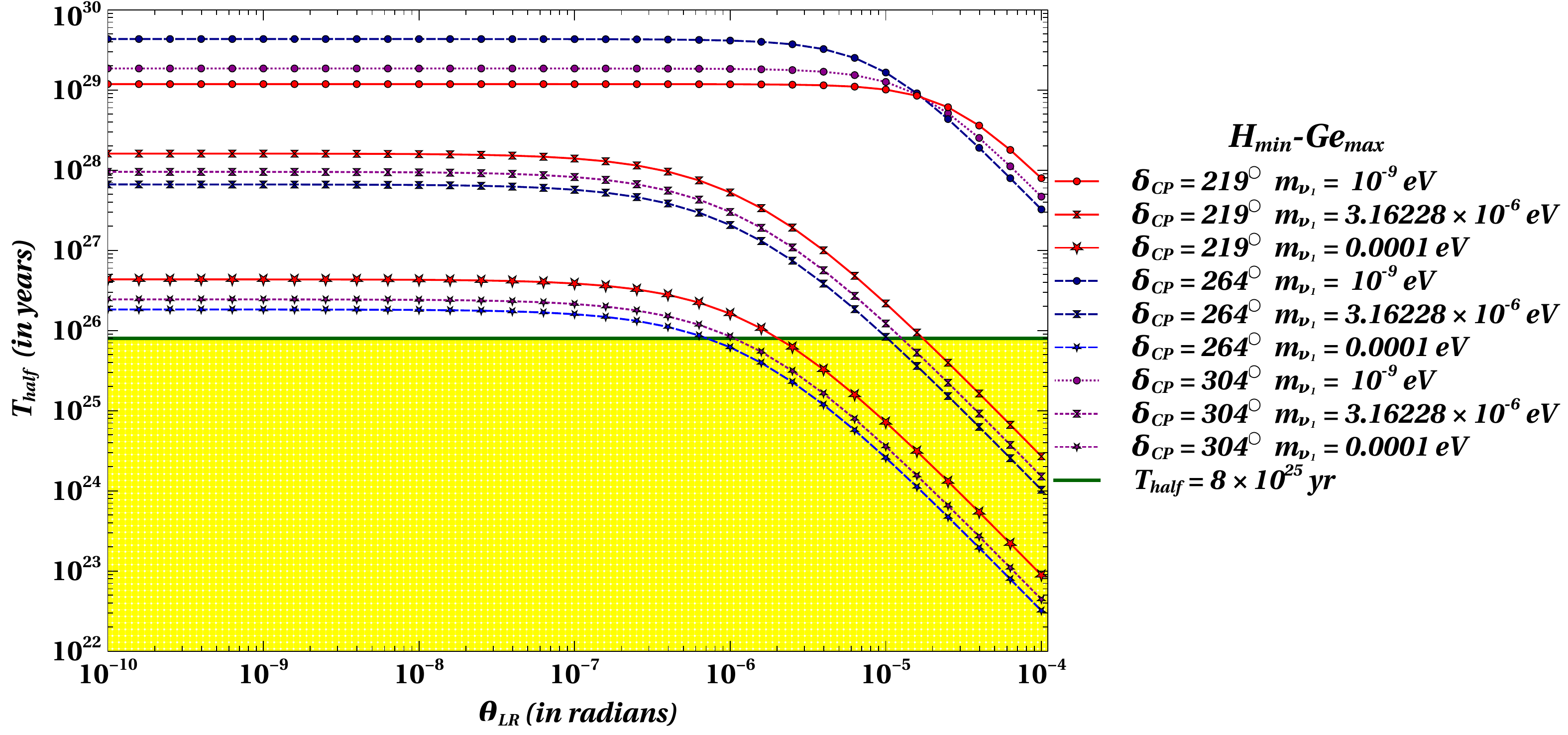}
\caption{Half-life of $0\nu \beta\beta$ process for ${}^{76}$Ge nucleus as a function of left-right charged gauge boson mixing for several fixed values of $m_{\nu_1}$ and $\delta_{CP}$.}
\label{fig:getheta}
\end{center}
\end{figure}

Fig.~\ref{fig:gemnu} shows the variation of $T_{1/2}^{0\nu}$ for ${}^{76}$Ge as a function of the lightest neutrino mass for several fixed values of $\theta_{LR}$ and $\delta_{CP}$. It is quite clear that the value of $T_{1/2}^{0\nu}$ decreases as the lightest neutrino mass and/or value of $\theta_{LR}$ increases. The variation with $m_{\nu_1}$ is quite evident from our earlier discussion. For the set of plots with $\theta_{LR}=10^{-8}$, the initial variation with $m_{\nu_1}$ is quite modest till $m_{\nu_1} \lesssim 10^{-7}$ eV. The major contribution here comes from the $\eta_1$ term with the half-life slowly decreasing with an increase in the lightest neutrino mass. At larger values of $m_{\nu_1}$, the light-heavy neutrino mixing increases significantly and the dominant contribution comes from the $\eta_{\lambda}$ term. The effect of increasing $\theta_{LR}$ can be understood as an artefact of an increase in the $\eta_{\eta}$ term which starts contributing quite significantly at $\theta_{LR} \gtrsim 10^{-6}$. Similar characteristics and dependence can be observed in Fig.~\ref{fig:getheta} where we have plotted the variation of $T_{1/2}^{0\nu}$ as a function of $\theta_{LR}$ for fixed values of $m_{\nu_1}$ and $\delta_{CP}$. As can be seen here, the value of $T_{1/2}^{0\nu}$ remain almost constant in the initial region of $\theta_{LR} \lesssim 10^{-6}$. In this region, the dominant contribution comes from $\eta_1$ and $\eta_{\lambda}$ and since the lightest neutrino mass is constant for each line, there is no variation in their value. In the region $\theta_{LR} \gtrsim 10^{-6}$, the $\eta_{\eta}$ term starts dominating and we see a sharp decrease in the value of $T_{1/2}^{0\nu}$ in this region.

\begin{figure}[h!]
\begin{center}
\includegraphics[scale=0.45]{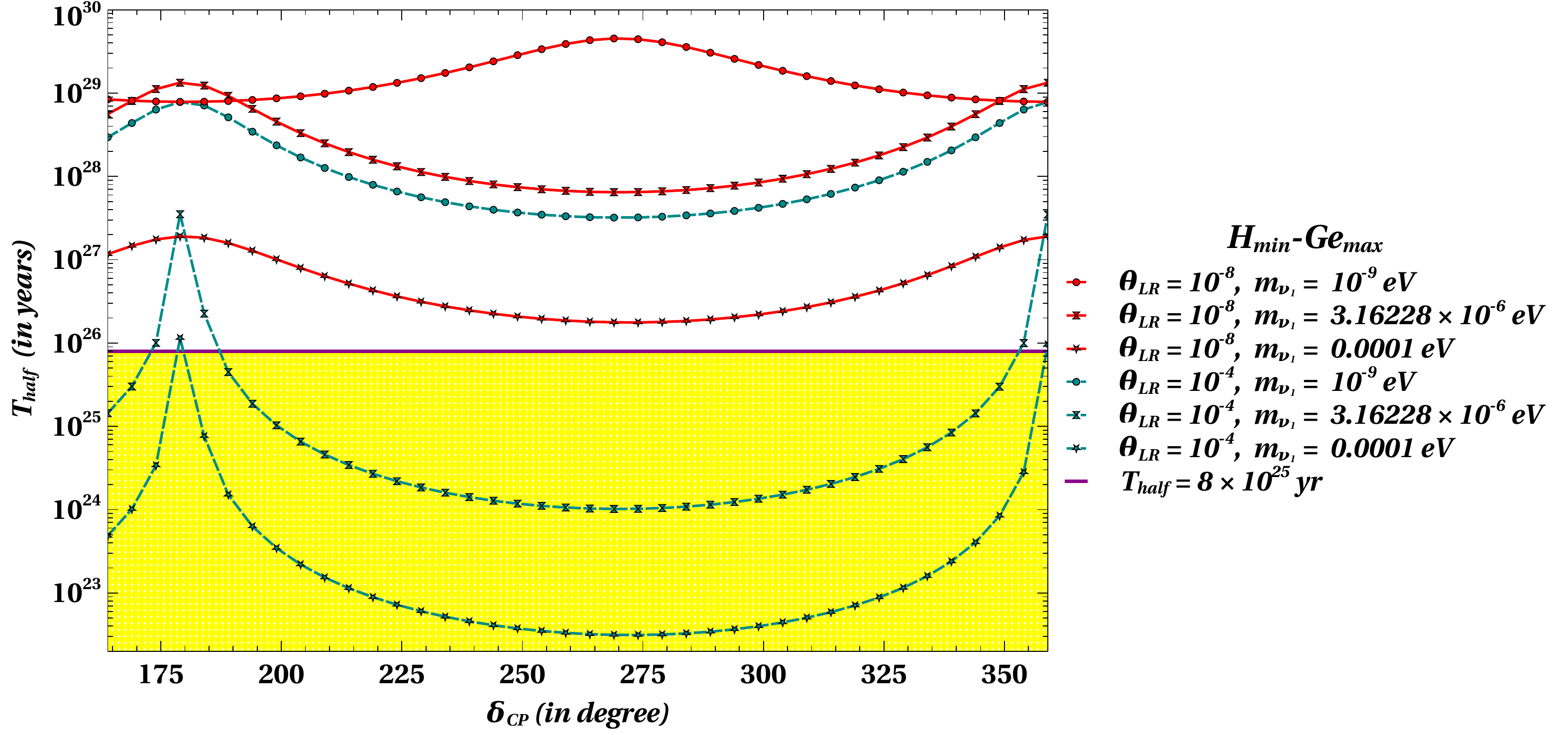}
\caption{Half-life of $0 \nu \beta \beta$ process for ${}^{76}$Ge nucleus as a function of the CP violating phase of the PMNS matrix for several fixed values of $\theta_{LR}$ and $m_{\nu_1}$.}
\label{fig:gedelta}
\end{center}
\end{figure}

Finally, the variation of $T_{1/2}^{0\nu}$ with $\delta_{CP}$ is given in Fig.~\ref{fig:gedelta}. This plot is quite interesting as it clearly shows the contribution of different $\eta$ terms in different regions of parameter space. The line corresponding to $\theta_{LR} = 10^{-8}$ and $m_{\nu_1} = 10^{-9}$~eV has a dominant contribution from $\eta_1$ term. As $\eta_1 = \frac{1}{m_e}\sum_i \mathbb{U}_{ei}^2 m_i$, its contribution in this case only depends on the matrix elements $\mathbb{U}_{ei}$ since $m_e$ and $m_i$ remains constant. The $\mathbb{U}_{13}$ element\footnote{The $\mathbb{U}_{13}$ element will be the same as the (1,3) element of the PMNS matrix.} is directly proportional to $e^{-i \delta_{CP}}$ and as a result one can approximately write $T_{1/2}^{0\nu} \sim \frac{1}{G_{01}^{0 \nu}\left| M_{\nu}^{0 \nu}\eta_1 \right| ^2}$. As $\delta_{CP}$ approaches $\ang{180}$ or $\ang{360}$, the value of $\eta_1$ goes through a maxima while it becomes a minima at $\delta_{CP}=\ang{270}$. The inverse of this behaviour is reflected in the $T_{1/2}^{0\nu}$ plot. For the other lines in this figure, they receive  dominant contributions from either $\eta_{\lambda}$ or $\eta_{\eta}$. Here as $\mathbb{U}_{13}$ decreases, the elements of $\mathbb{S}$ and $\mathbb{T}$ mixing matrices increase and hence their nature is opposite to the previous plot.  The nature of the plots for $^{136}$Xe nucleus is the same as the $^{76}$Ge nucleus except the fact that the parameters are more tightly constrained  owing to the more stringent experimental limit of the $0 \nu \beta \beta$ half-life for the $^{136}$Xe nucleus.

\begin{figure}[h!]
\begin{center}
\includegraphics[scale=0.95]{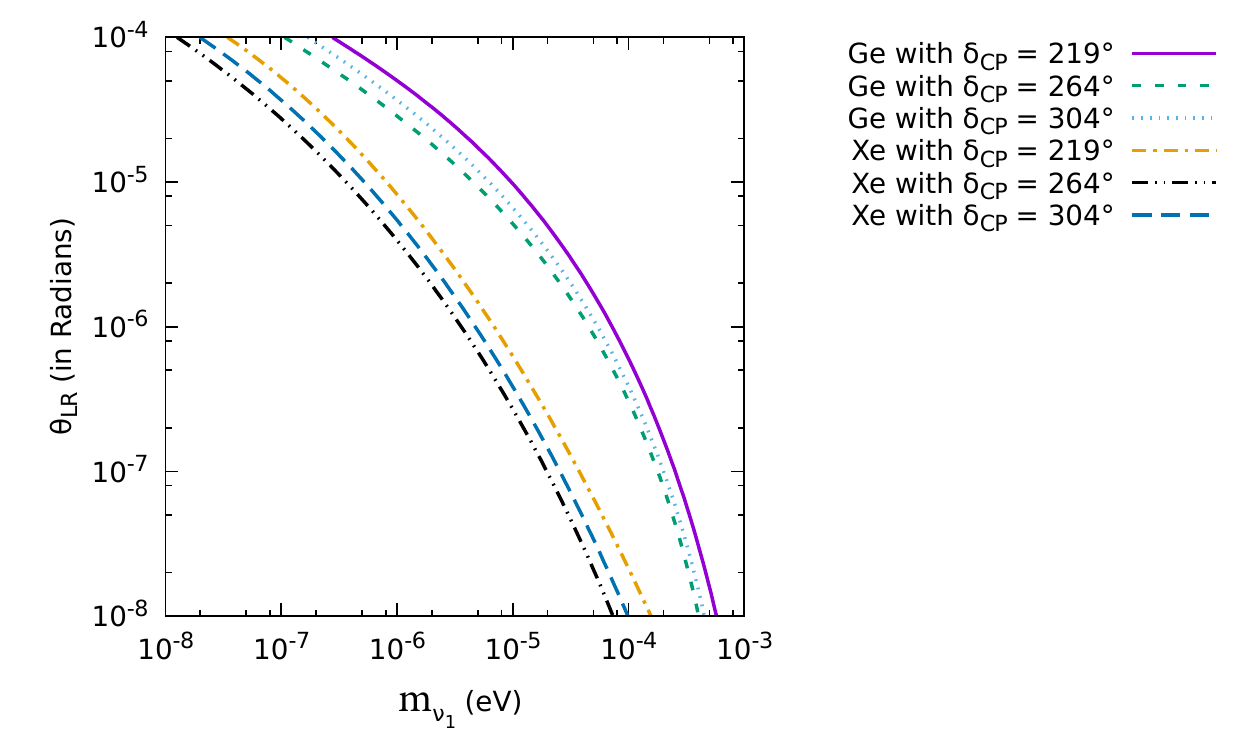}
\caption{Limits on the lightest neutrino mass and left-right $W$ gauge boson mixing.}
\label{fig:limit}
\end{center} 
\end{figure}

Fig.~\ref{fig:limit} shows the upper limits on the mass of the lightest neutrino in the this model as a function of $\theta_{LR}$ for fixed values of $\delta_{CP}$ for both $^{76}$Ge and $^{136}$Xe nucleus. As we have already discussed that the most stringent bound on the parameter space is obtained for a $\delta_{CP}$ of around $\ang{264}$, this fact is also reflected from this figure. As expected from the previous discussion, the upper limit on the lightest neutrino mass becomes much stronger for larger values of $\theta_{LR}$ and vice-versa. This occurs due to dominant  $\eta_{\eta}$ contribution for a large $\theta_{LR}$, leading to a tighter constraint on the lightest neutrino mass $m_{\nu_1}$.  Another observation is that the limits obtained for the $^{136}$Xe nucleus is much stronger than the $^{76}$Ge nucleus.

\section{Extended Seesaw Model and Analysis}
\label{s:excsaw}

This is another extension of SM, where the model contains light and heavy sterile neutrinos\footnote{We denote the gauge singlet neutrinos as sterile neutrino, as they are not charged under the SM gauge group.}, which can give large contribution in $0\nu \beta \beta$ process. Several studies \cite{Dekens:2020ttz,Alcaide:2017xoe,Vergados:2016hso,Cremonesi:2016jjo,DellOro:2016tmg,Peng:2015haa,Lisi:2015yma,Dev:2015vra,Faessler:2014kka,Pascoli:2013fiz,Huang:2013kma,Dev:2013vxa,Barry:2013xxa,Khan:2012zw,Wong:2012xdj,Chakrabortty:2012np,Mitra:2012qz,Vergados:2012xy,Chakrabortty:2012mh} have focused on sterile neutrinos with mass $>$ 100 MeV, and large contribution in the LNV process. Here, instead we consider some  of the sterile neutrino states in the $<$ 100 MeV mass range. We investigate the allowed model parameters, which satisfy a number of theoretical and experimental constraints. In doing so, we first consider a simplistic one generation scenario with one active neutrino, one light sterile neutrino $S_L$, and another heavy sterile neutrino $N_L$. Subsequently we extend our analysis with realistic three-generation case where the neutrino sector comprises of three active neutrinos along with the additional six sterile (three $S_{L}$ and three $N_{L}$) which are neutral under the SM gauge group. Below, we first review the model, and then discuss the contribution in $0\nu \beta \beta$ process. 
%(hence the name {\em sterile}). 

\subsection{Model}
\label{ss:Extcsawmodel}
%In this section, we give the most general set-up of Extended Seesaw model. We consider three generation of Standard Model (SM) neutrino $\nu_{L}$ along with heavy neutrino states $S_{L}$ and $N_{L}$.
The neutral lepton sector of the model contains three generation of SM neutrinos $\nu_L$ along with additional sterile neutrino states $S_L$ and $N_L$.  The  mass term of the neutrinos have the following form, 
\begin{equation}
\label{e:extcsawLag}
\mathcal{L}=-\frac{1}{2} \begin{pmatrix} \nu_{L} & S_{L} & N_{L} \end{pmatrix} \begin{pmatrix} 0 & 0 & M_{D}^{T} \cr 0 & \mu  & M_{S}^{T} \cr M_{D} & M_{S} & M_{R} \end{pmatrix} \begin{pmatrix} \nu_{L} \cr S_{L} \cr N_{L} \end{pmatrix}+~\rm{h.c}.
\end{equation}
We denote the neutral lepton mass matrix as $\mathcal{M}_{n}$, and hence  
\begin{equation}
\label{e:extcsawMat}
\mathcal{M}_{n} = \begin{pmatrix} 0 & 0 & M_{D}^{T} \cr 0 & \mu  & M_{S}^{T} \cr M_{D} & M_{S} & M_{R} \end{pmatrix}.
\end{equation}

We choose to work in a basis where the Majorana mass matrix $M_{R}$ of $N_{L}$ sterile is real. The term containing $\mu$ denotes the Majorana mass of the heavy  neutrino state $S_{L}$ with $\mu$ being a complex symmetric matrix. The Dirac mass matrix $M_{D}$ represents  the mixing between 
the SM  neutrino states $\nu_{L}$ and the heavy sterile neutrino states $N_{L}$; whereas, $M_{S}$ represents the mixing between the two sterile states $S_{L}$ and $N_{L}$. Throughout our analysis we consider the  matrices $M_{R}$ and $M_{S}$ are invertible. We also assume, that, the different sub-matrices follow the hierarchy, $M_{R} > M_{S} > M_{D} \gg \mu$. For seesaw approximation to be valid, the mixing matrices should satisfy $\mu < M_{S}^{T} M^{-1}_{R} M_{S}$, {\em i.e.},  $\mu < \mathcal{O}(\frac{M^2_S}{M_R})$, see~\cite{Mitra:2011qr} for details.

Contrary to the inverse seesaw~\cite{Mohapatra:1986aw,Mohapatra:1986bd,Wyler:1982dd,Witten:1985bz,Hewett:1988xc,Dev:2009aw,Blanchet:2010kw,Ilakovac:1994kj,Deppisch:2004fa,Arina:2008bb,Malinsky:2009df,Hirsch:2009ra}, Extended Seesaw model contains both the  heavy and small lepton number violation scales  $M_R$ and $\mu$ respectively. The SM neutrino masses  strongly depend  on the small lepton number violating scale $\mu$ and hence in the $\mu \to 0$ limit, the $\nu_L$ states become massless. The heavy Majorana neutrino contribution in $0\nu\beta\beta$ decay can be sizeable, even in the  $\mu \to 0$ limit. Hence, the contributions of the SM neutrinos and the heavy Majorana  neutrinos in $0\nu\beta\beta$ process are  completely decoupled from each other. Contribution from heavy sterile neutrinos for this model has been discussed in \cite{Mitra:2011qr}.

The neutrino mass matrix $\mathcal{M}_n$ can be diagonalized by a unitary transformation,
\begin{equation}
\label{e:diagonalization}
\mathcal{U}^{T} \mathcal{M}_n \mathcal{U} = \mathcal{M}_{n}^{d},
\end{equation}
where $\mathcal{U}$  as an expansion with order parameter $M_D/M_S$  has the following form~\cite{Mitra:2011qr}:

%\begin{scriptsize} 

\begin{flalign} \label{e:mixing1} 
\blds {\begin{pmatrix}(1-\frac{1}{2}M^{\dagger}_D(M^{-1}_S)^{\dagger}M^{-1}_SM_D) W_{\mu}  & M^{\dagger}_D (M^{-1}_S)^{\dagger}W_S  & M^{\dagger}_D M_R^{-1} W_N \cr -M^{-1}_S M_D W_{\mu} & 
(1-\frac{1}{2}M^{-1}_SM_DM^{\dagger}_D(M^{-1}_S)^{\dagger}\! -\!\frac{1}{2}M^{\dagger}_SM^{-2}_RM_S )W_S & M^{\dagger}_S M^{-1}_R W_N 
\cr
{M^T_S}^{-1}\! \mu M^{-1}_SM_D  W_{\mu} & -M^{-1}_R M_S W_S &  (1  -\frac{1}{2} M^{-1}_R M_S M^{\dagger}_S  M^{-1}_R )W_N \end{pmatrix}}.
\end{flalign}
%\end{scriptsize}

In the above, $W_{\mu}$, $W_{S}$ and $W_{N}$ are the three unitary matrices that diagonalize the block diagonal matrices 
\begin{equation}
\label{e:mass}
m_{\nu}  \sim   M_{D}^{T} {(M_{S}^{-1})}^{T} \mu {(M_{S})}^{-1} M_{D},\,~~
m_{s} \sim  -M_{S}^{T} {(M_{R})}^{-1} M_{S}, \,~~
m_{n}  \sim   M_{R}.
\end{equation}

The   matrix $m_{\nu}$ represents the light neutrino mass matrix, and $m_{s}$ and $m_{n}$ represent the heavy neutrino mass matrices. The hierarchy among the sub matrices ensures that $m_{n}$ and its eigenvalues give the heaviest sterile neutrinos in this model.  The other sterile neutrinos that originates from the diagonalization of $m_{s}$ can be relatively lighter, but they certainly should be heavier than the three active neutrinos $m_{\nu} <m_{s} < m_{n}$. In the subsequent sections, we explore the possibility of that the sterile states from $m_s$ are in the eV to MeV range, while the remaining sterile neutrino states $m_n$ are more than GeV. Before presenting a detailed analysis on $0\nu \beta \beta$, we first consider additional constraints coming from light neutrino mass measurement, non-unitarity and others. 

\subsection{Constraints}
\label{s:constraint}

Before delving into the analysis, we present a short descriptions of all constariants that has been applied in this model.

\begin{enumerate}[(a)]
\item  {\em Theoretical Constraints:} 
\begin{description}
\item[Hierarchy:]The different sub-matrices of Eq.~\ref{e:extcsawLag} should satisfy the hierarchy  \begin{itemize}
\item $M_R> M_S> M_D \gg \mu$,\\
and
\item $M_{S}^{T}M_{R}^{-1}M_{S} > \mu$; {\em i.e.}, $m_{s} > \mu$ (from Eq.~\ref{e:mass}), for one generation this will be $M_{S}^{2}/M_{R} > \mu$. This limit also defines the region where seesaw approximation is valid~\cite{Mitra:2011qr}.\end{itemize}
\item[Unitarity:]The mass matrix being symmetric, the diagonalization matrix given in Eq.~\ref{e:diagonalization} should be orthogonal or unitary, {\em i.e.}, we should have the relation $\mathcal{U}^{\dagger}\mathcal{U} = \mathcal{U}\mathcal{U}^{\dagger} = \mathbb{{\bf I}}$; but working with the seesaw approximation up to a certain order and also having low scale sterile, the elements of $U = \mathcal{U}\mathcal{U}^{\dagger}$ will not be identity matrix rather those elements will be $\mathbb{{\bf I}} \pm \delta$, where $\delta$ is the tolerance of every single elements of $U$ to get a viable parameter space for lightest sterile\footnote{See analysis section.} in this model.
So, to zero in on the allowed parameter space of eV to MeV sterile in this model, we have to constrain the parameter space, setting some cut-off values on the both diagonal and non-diagonal elements of $U$. In short, we allow some error bar on the diagonal elements on $U$ around unity and for non-diagonal elements the required error bar will be around zero. Depending on the choice of parameter space, the error bar may differ for diagonal and non-diagonal elements. We have generally taken the the maximum constraints on the deviation which provides us the desired allowed parameter space. 

\end{description}
\item {\em Experimental Constraint:}
\begin{description}
\item[Mass of active neutrino:]We consider the constraint on the sum of active neutrino masses from Planck cosmological data~\cite{Ade:2015xua}, {\em i.e.}, at $95\%$ C.L. the sum on the masses of active neutrinos will be less than $\sum m_{\nu} < 0.194$ eV. In the analysis of one-generation, this bound simply manifests as the upper bound on the mass of single active neutrino. We implement the constraints on mixing angles and on the mass square differences among three active neutrinos from neutrino oscillation data in the three-generation case~\cite{Esteban:2020cvm,deSalas:2020pgw} in case of Normal hierarchy.
\item[Constraints from $0\nu \beta\beta$ limit:]The limit on the $T_{1/2}^{0\nu}$ from the KamLAND-Zen~\cite{KamLAND-Zen:2016pfg} severely constrains the parameter space for eV/MeV sterile of this model, see Sec.~\ref{s:0-nu-beta-decay}.

\item[Daya Bay experiment:]The Daya Bay reactor anti neutrino experiment shows a large exclusion region between $2 \times 10^{-4} < \Delta m_{s1}^{2} < 0.3 ~{\rm eV}^{2}$ as function of $\sin ^{2} 2\theta_{1s}$~\cite{An:2016luf} at $95\%$ CL, where $\Delta m_{s1}^{2}$ is the mass-square difference between extra sterile and electron-neutrino ($\nu_{e}$) and $\theta_{1s}$ is the angle of active-sterile mixing. This result will further constrain the allowed parameter space for eV sterile. 
\end{description}
\end{enumerate}

\subsection{$0\nu\beta\beta$ decay: sterile neutrino contributions}\label{s:0-nu-beta-decay}

In this section, we outline the contributions of sterile neutrinos having Majorana masses in $0\nu \beta\beta$ decay.  
  
The  half-life of $0\nu \beta\beta$ is written as~\cite{Mitra:2011qr, Kovalenko:2009td} 
\begin{equation}
\label{e:Thalf2}
\frac{1}{T_{1/2}^{0\nu}}=K_{0\nu} \left| \Theta_{ej}^2 \frac{ \mu_j}{\langle p^{2} \rangle-\mu^{2}_{j}}  \right|^{2},
\end{equation}
where $j$ represents the number of light neutrino states and the additional heavy neutrino states.  The parameters $\mu_{j}$ and $\Theta_{ej}$  represent the masses of the neutrino states and the mixing with SM neutrinos respectively. In the above,  $K_{0\nu} = G_{0\nu} (\mathcal{M}_{N} m_{p})^{2}$ and  $\langle p^{2} \rangle\equiv -m_{e} m_{p} \frac{\mathcal{M}_{N}}{\mathcal{M}_{\nu}}$. The reference mass scales are considered as electron ($m_{e}$) and proton ($m_{p}$) masses, $M_{\nu}$ and $M_{N}$ are the NMEs for exchange of light and heavy neutrinos respectively. The values of NME and phase space factor $G_{0\nu}$ have been taken from Ref.~\cite{Meroni:2012qf}. Below, we classify the sterile neutrino contributions according to its mass scale. 

Other than the contributions of the SM neutrinos, the sterile neutrino states $S_{k}$ and $N_{k}\, ~(k=1,2,3;~ {\rm in~our~case})$ can also contribute in the $0\nu\beta\beta$ process. Evidently, we have two extra contributions apart from the SM one. 

\begin{enumerate}[(a)]
%\begin{description}
\item The heaviest states $N_{k}$ have a mass range $m_{{n}_{k}}$ $>$ 100-200 MeV and give a contribution in $0\nu\beta\beta$ as
\begin{equation}
\label{e:0nubNmcontbtn}
{A}_{N} \sim  \frac{{V}^2_{eN_k}}{m_{{n}_{k}}},
\end{equation}
where $A_N$ represents the amplitude of this process, and $V_{eN_{k}}$ is the mixing of the $N_{k}$ states with the active neutrinos. Using $V_{eN}={M^{\dagger}_{D}}{M^{-1}_{R}} W_{N}$, this can be simplified as 
\begin{equation}
\label{e:amplN}
{A}_{N} \sim  (M_{D}^{T} M_{R}^{-3} M_{D})_{ee}.
\end{equation}
\item The other sterile neutrino states $S_{k}$ give  contributions proportional to 
\begin{equation}
\label{e:0nubSmcontbtn}
{A}_{S} \sim  \frac{{V}^2_{eS_{k}}}{m_{{s}_{k}}},
\end{equation}
for the mass range $m_{{s}_{k}}$ $>$ 100-200 MeV, whereas 
\begin{equation}
\label{e:amplS}
{A}_{S} \sim  \frac{{V}^2_{eS_k}m_{{s}_{k}}}{\langle p^{2} \rangle},
\end{equation}
when sterile mass is light. We use the compact expression for the amplitude, that also take into account $\langle p^{2} \rangle  \simeq m_{s_k}^{2} \simeq 100 \, \rm{ MeV}^2$. 
\begin{equation}
\label{e:amplS_small}
{A}_{S} \sim  \frac{{V}^2_{eS_k} m_{s_k}}{\langle p^{2} \rangle -m_{{s}_{k}}^{2}}.
\end{equation}
%\end{description}
\end{enumerate}

The value of $m_{{n}_{k}}$ in our analysis is $10^{5}$ GeV for one-generation case giving rise to active-sterile mixing as $M_{D}/M_{R} \sim 10^{-10} - 10^{-7}$, whereas $m_{{n}_{k}}$s are to be of the order of 50 to 500 GeV in the three-generation one having ${V}^2_{eN_k}$ as $\sim 10^{-8}$ to $10^{-7}$. So, the sterile neutrinos $N_{k}$ being heavy contributes negligibly and hence we do not consider its contribution. The half-life of $0\nu \beta\beta$ is thus,

\begin{equation}
\label{e:Thalf1}
\frac{1}{T_{1/2}^{0\nu}}=K_{0\nu} \left| \frac{U_{ej}^2  m_{{\nu}_j}}{\langle p^{2} \rangle- m_{{\nu}_j}^{2}}  +  \frac{{V}^2_{eS_k} m_{s_k}}{\langle p^{2} \rangle -m_{{s}_{k}}^{2}}\right|^{2},
\end{equation}
where $j$ represents the number index of three light neutrino states whereas $k$ denotes the number index of comparatively light sterile states $S_{L}$. 
This is to note that, the lighter $S_{L}$ states (eV-MeV) can have substantial active-sterile mixing $V_{eS} = {M_{D}^{\dagger}}{({M_{S}^{-1})}^{\dagger}} W_{S}$.

To get the essence of all theoretical and experimental constraints properly, first we consider only one generation for all states, so the mixing matrix given in Eq.~\ref{e:mixing1} will be of the order of $3\times 3$ having $M_{D}, M_{R}, M_{S}, \mu$ as simply numbers.

\subsection{Analysis:}
\label{ss:res_excsaw}

\begin{figure}[hbt!]
\begin{center}
\subfloat[\label{sf:pMeVonegen}]{
\includegraphics[scale=0.57]{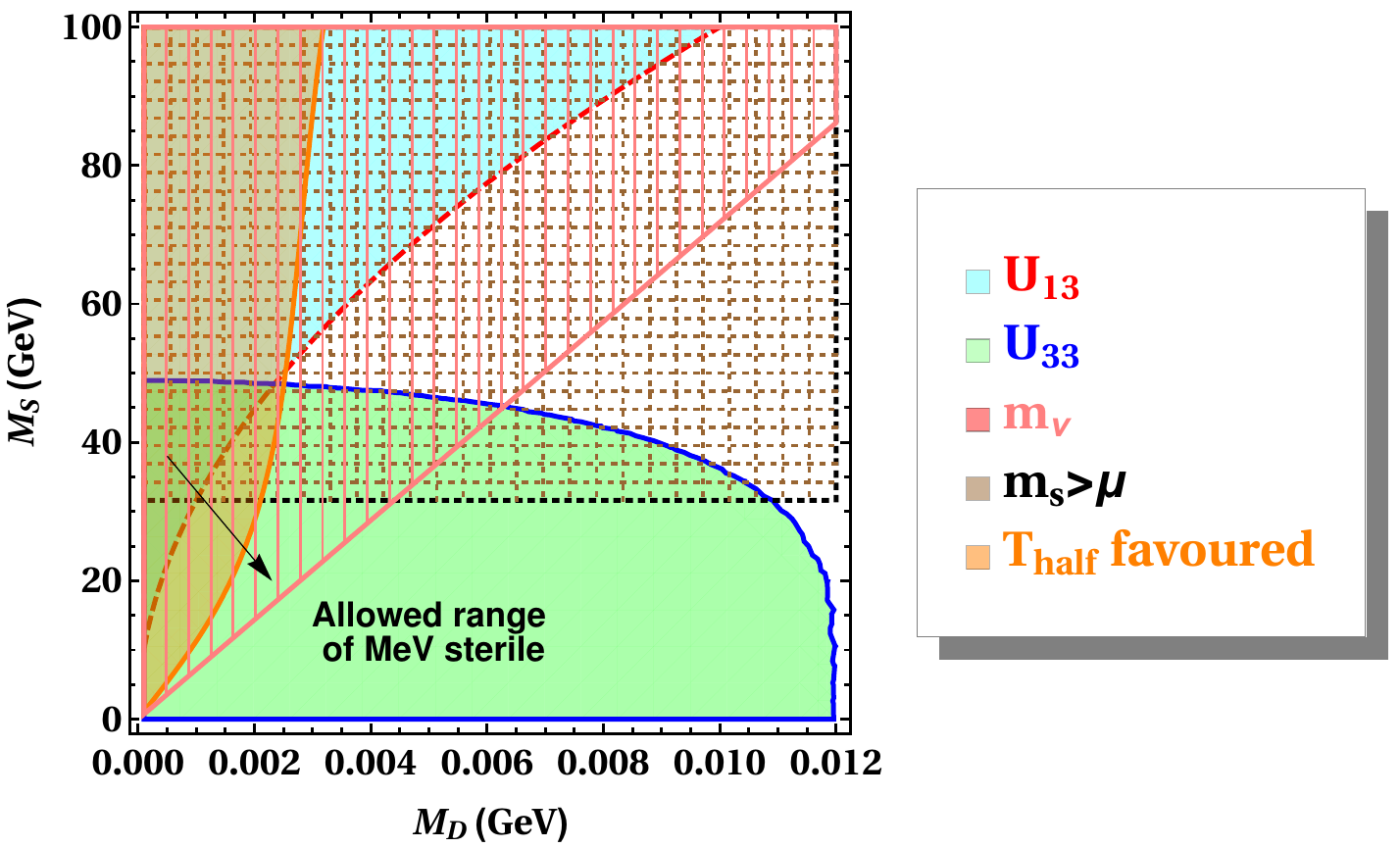}}
~~~~
\subfloat[\label{sf:peVonegen}]{
\includegraphics[scale=0.57]{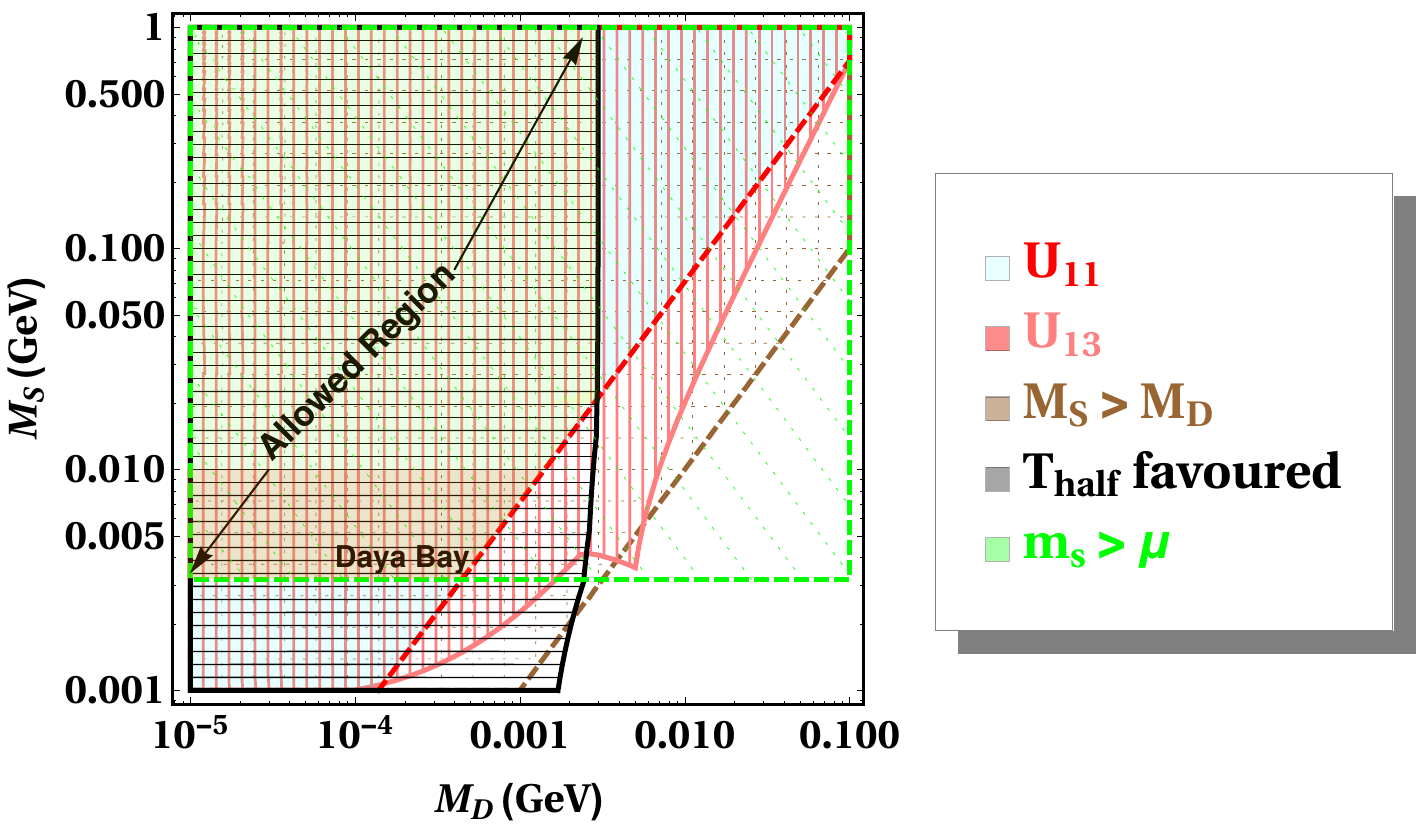}} 
\caption{Allowed parameter space of light sterile neutrino in (a) MeV and in (b) eV range as function of the model parameters $M_{D}$ and $M_{S}$ in Extended Seesaw scheme. The regions have been obtained from theoretical constraints, light neutrino measurements and  $0\nu\beta\beta$ results. The parameter $\mu$ has been set to $10^{-2}$ GeV and $10^{-10}$ GeV for MeV and eV range respectively. In both cases, $M_{R} = 10^{5}$ GeV. The orange shaded region of (b) will be constrained further from reactor anti-neutrino Daya Bay experiment~\cite{An:2016luf} that will be elaborated in Fig.~\ref{Daya_Bay_constriant}.}
\label{f:sterilevalue}
\end{center}
\end{figure}

\subsubsection{One Generation}
\label{sss:onegen}

In this section, we provide a detailed analysis and results of the allowed range of MeV/eV sterile neutrino. In Fig.~\ref{f:sterilevalue}, the plot in the left panel (Fig.~\ref{sf:pMeVonegen}) shows the allowed region for MeV sterile and its contribution to $0\nu\beta\beta$ in $M_{D}- M_{S}$ plane for one-generation scenario. The matrix $\mathcal{M}_{n}$ in Eq.~\ref{e:extcsawMat}, is $3 \times 3$ instead of being $9 \times 9$ in this case. The square boxes in the index box of this figure (Fig.~\ref{sf:pMeVonegen}) shows the color of allowed regions in agreement with different constraints and the respective texts are written in the same color of the border of that region. The cyan colored region enclosed by the red-dashed curve in Fig.~\ref{sf:pMeVonegen} shows the region allowed by the off-diagonal element $U_{13}$, where $U_{13}$ is the (1,3) element of $\mathcal{U}^{\dagger}\mathcal{U}$ with $\mathcal{U}$ being the diagonalization matrix (Eq.~\ref{e:diagonalization}). Here, we consider the constraint as $|U_{13}| < 10^{-8}$, {\em i.e.}, $U_{13}$ is almost vanishing. The lower green region enclosed by blue solid line presents the region allowed by the diagonal element $U_{33}$. The constraints due to other matrix elements of $U$  coming from the condition of diagonalization matrix $\mathcal{U}$ being unitary are less stringent and are simply allowed by the final overlapped zone. In that final allowed region, the numerical value of off-diagonal element $U_{13} ~{\rm is} \sim 10^{-9}$ and that of diagonal element $U_{33}$ is $\sim (1 + 10^{-16})$. The region covered by pink-colored straight lines shows the mass of light neutrino range $0 < m_{\nu} < 0.194$ eV. The brown rectangle region enclosed from below by the black dashed line (near $M_{S} = 32$ GeV) represents the constraint $m_{s} = M_{S}^{2}/M_{R} > \mu$ marking the area where the seesaw approximation is valid. The extreme left almost vertical orange-colored region enclosed by the solid orange bow-type curve shows the region in agreement with the contribution of $0\nu \beta\beta$, where $T_{1/2}^{0\nu} > 1.07 \times 10^{26}$ yr~\cite{KamLAND-Zen:2016pfg}. In obtaining this allowed parameter space, we have considered both the light neutrino and sterile neutrino contribution, see Eq.~\ref{e:Thalf1}. The values of the NMEs that we have considered in this analysis are $M_{\nu} = 2.29$ and $M_{N} = 163.5$~\cite{Meroni:2012qf}.

The overlapped region in Fig.~\ref{sf:pMeVonegen} enclosed by dashed black straight line from below and  solid blue line from above with  32 GeV $< M_{S} <$ 49 GeV and red-dashed line from the right is the final allowed range for MeV sterile in Extended Seesaw model, with the value of $M_{D}$ $\leq$ 0.00011 GeV for $M_{S} \sim$ 32 GeV and with $M_{D}$ value up to $\sim$ 0.002356 GeV for $M_{S} = 49$ GeV. 
The allowed mass range of $m_{s}$  %from both theoretical point of view and from $0\nu \beta\beta$ 
is 10 MeV $< m_{s} <$ 24 MeV. The mass of the active neutrino in that region is $m_{\nu} \leq 10^{-2}$ eV.
%For $M_{S} = 32$ GeV, the maximum allowed value of $M_{D}$ is 0.001042 GeV and for $M_{S} = 49$ GeV the corresponding $M_{D}$ value is $\sim$ 0.002356 GeV.

The plot in the right panel (Fig.~\ref{sf:peVonegen}) of Fig.~\ref{f:sterilevalue} shows the allowed region for eV sterile in the Extended Seesaw model and its contribution to $0\nu\beta\beta$ decay in $M_{D}- M_{S}$ plane. The inclusion of sterile neutrinos whether being heavy or light has its effect on the unitarity of PMNS matrix~\cite{Blennow:2016jkn}. The PMNS matrix encoding the non-unitarity effect due to the mixings of active-sterile neutrinos is given by~\cite{Blennow:2016jkn}
\begin{equation}
\label{e:nonunitarity}
\mathcal{N}= (I - \alpha)\mathcal{U}^{\prime},
\end{equation}
where, $\mathcal{U}^{\prime}$ is equivalent to standard PMNS matrix which is also a unitary matrix having small deviation proportional $\alpha$. Clearly from Eq.~\ref{e:mixing1} we can see that $\alpha$ being theoretically same for all elements, is given by $1/2(M_{D}^{2}/M_{S}^{2})$ in Extended Seesaw model, whereas $\mathcal{U}^{\prime}$ is equivalent to $W_{\mu}$ of the same equation (\ref{e:mixing1}). The general form of mixing terms is usually given by the ratio between the mass-scales of light neutrinos and the sterile neutrinos. Therefore, in case of light sterile ($\sim$ eV), the active neutrino and the light sterile mass-scales being very close the effect of mixing can not be ignored. 
%In case of sterile being light and having the mass of the order of eV, the masses of active neutrino and light steriles are very close, therefore, the effect of mixing can not be ignored as the general form of  mixing terms is usually given by the ratio between the masses of light neutrinos and sterile neutrinos. 
So the light sterile has strong impact on the deviation of PMNS matrix from being unitary. In the Ref.~\cite{Blennow:2016jkn}, the constraint on $\alpha$ is given for different mass values of sterile neutrinos. For the mass-square difference of sterile and active neutrinos in the ${\rm eV}^{2}$ regime,  at $95\%$ CL, the bound is given by $\alpha < 10^{-2}$. So, in case of $\mathcal{N}^{\dagger}\mathcal{N}$ and for one-generation this bound is manifested as of the order of $\sim 10^{-4}$ as the deviation from unity $({\rm for~ the~element}~U_{11})$.

%\begin{wrapfigure}{l}{0 pt}%0.4\textwidth}
\begin{wrapfigure}[22]{r}{0.45\textwidth}
%\vspace{-22pt}
\centering
 \includegraphics[scale=0.5]{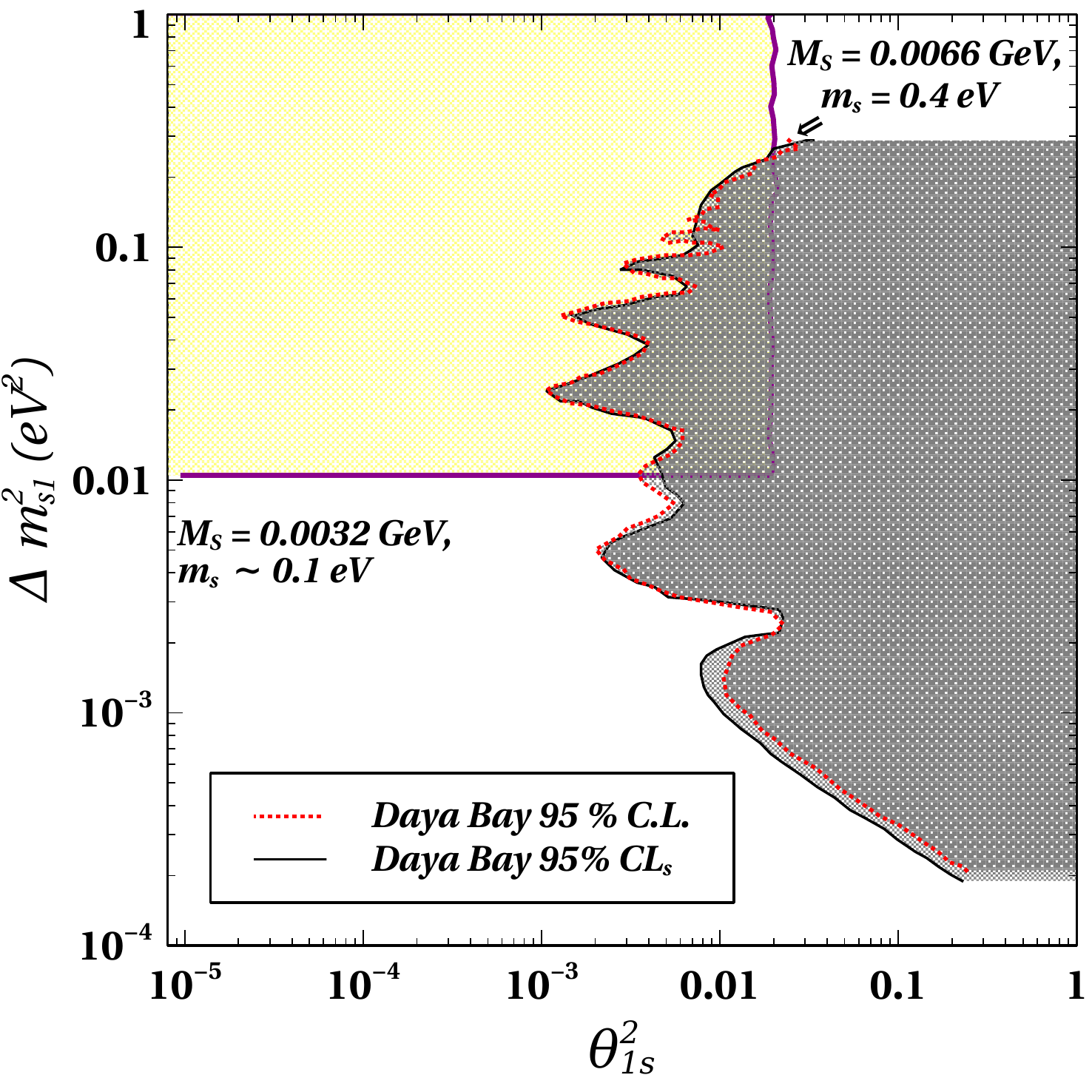}
%\vspace{-22pt}
\caption{\small Constraining the region of eV sterile from Daya Bay reactor neutrino experiment. The gray-colored region from right side shows exclusion region from Daya Bay results, whereas, the yellow colored region shows the zone allowed for eV sterile.}
\label{Daya_Bay_constriant}
%\vspace{-25pt} 
\end{wrapfigure}
%\lipsum[1-6]

Therefore, in Fig.~\ref{sf:peVonegen}, the constraint on $U_{11}$ is taken as $|U_{11} -1| < 10^{-4}$. Since, the matrix $\mathcal{U}_{22}$ in Eq.~\ref{e:mixing1} also contains a term like $(1 - 1/2(M_{D}^{2}/M_{S}^{2}) - 1/2(M_{S}^{2}/M_{R}^{2}))$ and $M_{R}$ being heavy this term can be effectively written as $(1 - 1/2(M_{D}^{2}/M_{S}^{2}))$; so the constraint on $U_{22}$ will be the same as that of $U_{11}$. The cyan-colored region surrounded by the red-dashed curve show the allowed region for an eV sterile from the limit applied on $U_{11}$. The pink colored region shows the allowed region for eV sterile as $|U_{13}| < 10^{-8}$. The brown-colored region shows the corresponding allowed range in agreement with the hierarchy of the model parameters $M_{S} > M_{D}$. The black-colored region depicts the allowed region from $0\nu\beta\beta$ decay~\cite{KamLAND-Zen:2016pfg} where contributions of eV sterile and active neutrinos have been considered. The green-colored region of oblique line shows the validity range of seesaw approximation. The final allowed region is given by the (orange + yellow) colored shaded region. The allowed region is enclosed by green line from below (at $M_{S} =$  0.0032 GeV, $M_{D} \sim 5 \times 10^{-4}$ to 0.002 GeV), by tilted red-dashed line from  lower-right ($(M_{D}, M_{S}) \sim 5 \times 10^{-4}$, 0.0032 GeV up to $(M_{D},M_{S}) \sim$  (0.002, 0.02) GeV) and finally by almost vertical black line from the right side. The region can be further extended leftwards and upwards by lowering the value of $M_{D}$ and increasing the value of $M_{S}$ respectively.
The other constraints such as unitarity constraints from other elements of $U$ and limit from neutrino mass is allowed by the final overlapped region. The lowest allowed value of model parameter $M_{S}$ is 0.0032 GeV at $M_{D} =10^{-5}$ GeV. Evidently, the lowest value of sterile neutrino from the allowed region showed in this figure is about 0.1024 eV and $m_{\nu} \leq 10^{-3}$ eV. In this region the $U_{11} \sim U_{22} \sim 1 + 10^{-7}$ approximately and $U_{13} \sim 10^{-9}$. The values of NMEs are the same as that of the Fig.~\ref{sf:pMeVonegen}.

Since, in Extended Seesaw, we have eV sterile starting from $\sim$ 0.1 eV, the region can further be constrained from reactor anti-neutrino experiments, such as, Daya Bay. Evidently, the orange shaded region of Fig.~\ref{sf:peVonegen} or a part of it, where we have sterile $\sim \mathcal{O}$(eV) can be probed again from such results. We represent the mentioned region of Fig.~\ref{sf:peVonegen} of $M_{S} - M_{D}$ plane in $\Delta m_{s1}^{2} ({\rm eV}^{2}) - \theta_{1s}^{2}$ in the Fig.~\ref{Daya_Bay_constriant}. The filled-in yellow box covered by the magenta color line corresponds to the aforementioned region of eV sterile of Fig.~\ref{sf:peVonegen}. The lower line corresponds to the value $M_{S} = 0.0032$ GeV where as upper line corresponds to $M_{S} = 0.01$ GeV. 
 
The region in Fig.~\ref{sf:peVonegen}, below the green line and right side of the red-dashed line being completely ruled out from our model parameters (Fig.~\ref{sf:peVonegen}) represents the white regions of Fig.~\ref{Daya_Bay_constriant}.
 The dotted red and black solid line represent the Daya Bay experimental constraints on low-scale sterile in $\Delta m_{s1}^{2} ({\rm eV}^{2}) - \theta_{1s}^{2}$ plane~\cite{An:2016luf}. The gray-colored region shows the zone that are not allowed by this experimental data. The overlapped region of this adjacent figure is ruled out from the experimental result. We can see that Daya Bay results exclude some region from the left side (giving constraint on $M_{D}$) but still allow all the values of sterile neutrino from 0.1 eV to 0.4 eV. The mass of active neutrino in the remaining allowed zone is $m_{\nu} \leq 10^{-4}$ eV.

In passing, we would like to comment on the active-sterile mixing value which  is getting constrained from Daya Bay data. The $\theta_{1s}^{2}$ is actually ${(M_{D}/M_{S})}^{2}$. Also, from  unitarity~\cite{Blennow:2016jkn}, we have $\frac{1}{2}{(M_{D}/M_{S})}^{2} < 10^{-2}$. Fig.~\ref{Daya_Bay_constriant} shows for some $M_{S},M_{D}$ values $\theta_{1s}^{2} < 10^{-3}$ giving slightly more stringent bound on the mixing compared to that of coming from unitarity in our model set-up.

\subsubsection{Three Generation}
\label{sss:3gen}
In Section \ref{sss:onegen}, we have discussed different constraints from neutrino mass, half-life of $0\nu \beta \beta$ decay, unitarity and from the validation of seesaw approximation for one generation realisation of Extended Seesaw model. In this section, we are extending the analysis for three-generation case which is more realistic than the previous scenario. In addition to the bounds from $0\nu \beta \beta$, other experimental and theoretical constraints, we also satisfy  neutrino oscillation data. In particular, we consider that,
\begin{itemize}
\item upper bound on the sum of all three active neutrinos is constrained from cosmology,
$\sum_i m_{\nu_i} < 0.194$ eV at $2\sigma$ C.L. \cite{Ade:2015xua},
\item two mass squared differences $6.93<\dfrac{\Delta m^2_{21}}
{10^{-5}}\,{\text{eV}^2} < 7.97$ and $2.37<\dfrac{\Delta m^2_{31}}
{10^{-3}}\,{\text{eV}^2} < 2.63$ vary in the $3\sigma$ range \cite{Esteban:2020cvm},
\item $3 \sigma$ range \cite{Esteban:2020cvm} of the three mixing angles $30^{\circ}<\,\theta_{12}\,<36.51^{\circ}$,
$37.99^{\circ}<\,\theta_{23}\,<51.71^{\circ}$ and
$7.82^{\circ}<\,\theta_{13}\,<9.02^{\circ}$.
\end{itemize}  
In the present set-up, contribution to the $0\nu\beta\beta$ can come from light active neutrinos ($m_{\nu_i}, i=1,2,3$), additional eV to MeV scale sterile neutrinos ($m_{s_i}, i= 1, 2, 3$) and heavy GeV scale neutrinos ($m_{n_i}, i= 1, 2, 3$). As shown in Sec.~\ref{s:constraint}, the contribution of heaviest sterile neutrinos to 
$0 \nu \beta \beta$  is evidently suppressed. Here, the neutrino mass matrix is $9\times9$, but we are working in the seesaw approximation regime which gives three sets of $3\times3$ matrices namely $m_{\nu}$, $m_{s}$ and $m_{n}$. After diagonalization of each $3\time 3$ block of Eq.~\ref{e:mass} individually, we check the unitarity constraints as described in Sec.~\ref{s:constraint} and in Sec.~\ref{sss:onegen}. The matrices which diagonalize each blocked-matrices combine to form a $9\times 9$ matrix (Eq.~\ref{e:mixing1}) and we impose constraints on the unit matrix ($U$, see Sec.~\ref{sss:onegen}) with the absolute variation of each elements by $\pm 10^{-2}$.
%diagonal elements
%around 1 with a variation of $\pm 10^{-2}$ and for the off-diagonal
%elements same variation with respect to 0. 
This $\pm 10^{-2}$ variation manifestly impose constrains on the ratio
$M_{D}/M_{S}$.
% and there is also bound on the ratio of these parameters. 
We have checked that the error bar is consistent with the experimental bound \cite{Blennow:2016jkn} that arises due to the non-unitarity effect for eV-keV scale sterile neutrino. A detailed description of different conditions provided in Sec.~\ref{s:constraint} to constraint the parameter space have been thoroughly followed in the present analysis. 

As discussed earlier, after using the seesaw approximation, we get three different $3\times 3$ matrices, which are $m_n, m_s, m_\nu$. Among them one corresponds to the mass matrix for the three active neutrinos (denoted by $m_{\nu}$), other two correspond to the mass matrices for the three relatively light sterile neutrinos (denoted by $m_s$) and the three heavy neutrinos which are in GeV scale (denoted by $m_n$). From Eq.~\ref{e:mass}, we can see the expressions of $m_{\nu}$, $m_{s}$ and $m_{n}$ depend on the matrices $M_{D}$, $M_{R}$, $M_{S}$ and $\mu$ whose elements are the free input parameters in the extended seesaw scenario. We choose the model parameters in our framework in a way so that we can accommodate eV to MeV scale sterile neutrino. 
%To that end, we have chosen the values of the $M_{D}$, $M_{R}$, $M_{S}$ and $\mu$ elements in such a way so that we get eV scale sterile neutrinos. 
Before proceeding, we consider few assumptions which include $M_{D}$, $M_{S}$, $M_{R}$ as the real diagonal matrices and $\mu$ as the complex symmetric matrix {\it i.e.} $\mu^{T} = \mu$ and $\mu^{*} \neq \mu$. In this work, we have focused on the normal hierarchy of the neutrino masses as illustrative example.
 In order to satisfy neutrino oscillation constraints and to obtain sterile neutrinos in eV to MeV scale, we have varied the model parameters in the following range (in GeV),
\begin{eqnarray}
10^{-5} & \le M_{D\, ii} &\le 10^{-1},  \nonumber \\ 
10^{-3} & \le M_{S\, ii} &\le 10^{-1}, \nonumber \\
50 & \le M_{R\, ii} &\le 500, \nonumber \\
10^{-11} & \le \mu^{R, I}_{ij} &\le 10^{-8}, \nonumber \\
\label{parameter-range}
\end{eqnarray}      
where $i, j$ can vary from 1 to 3. Below we show the allowed model parameters as well as correlation between different observables for this model as scatter plots. 
 
\begin{figure}[h!]
\centering
\includegraphics[angle=0,height=7cm,width=8cm]{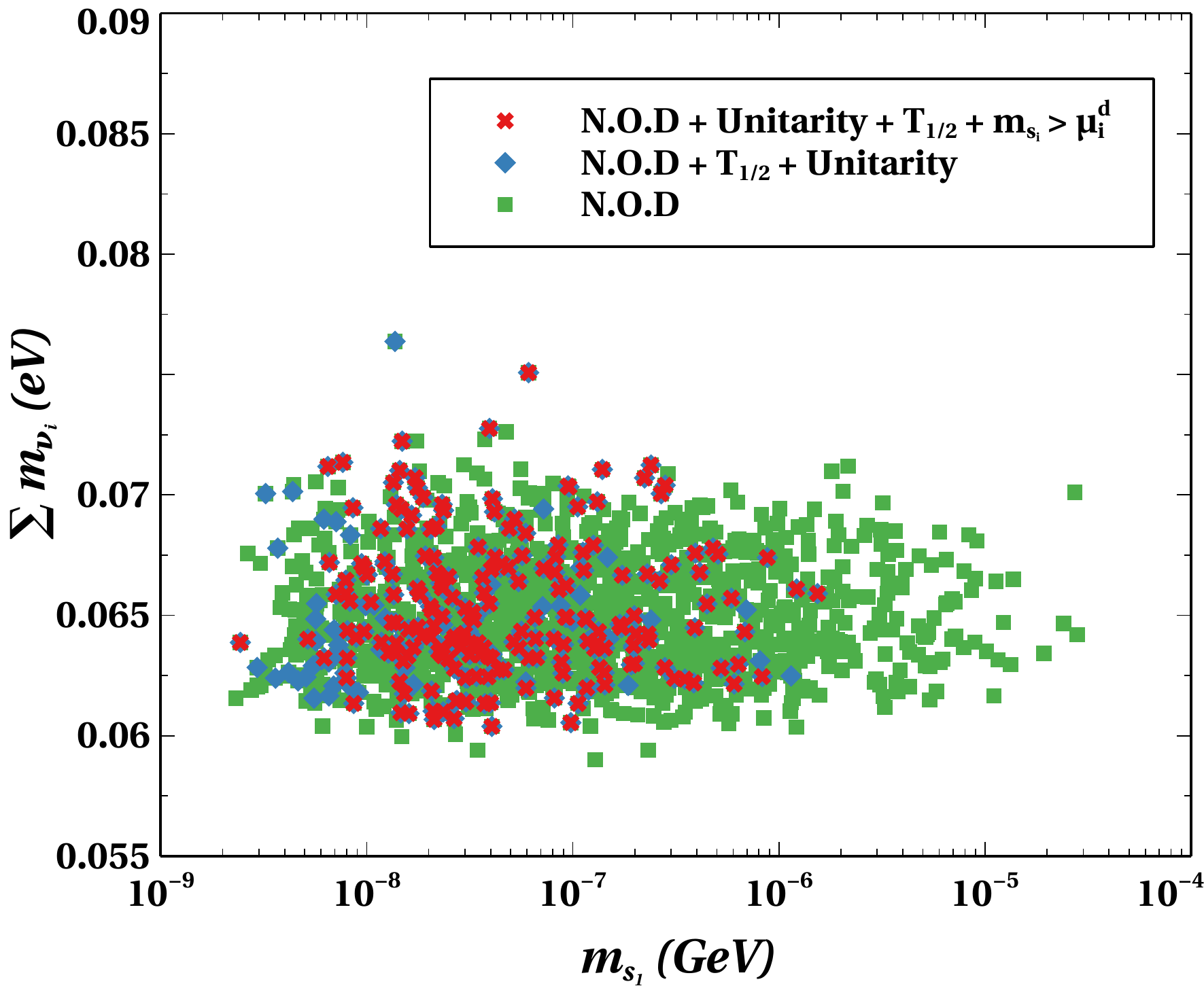}
\includegraphics[angle=0,height=7cm,width=8cm]{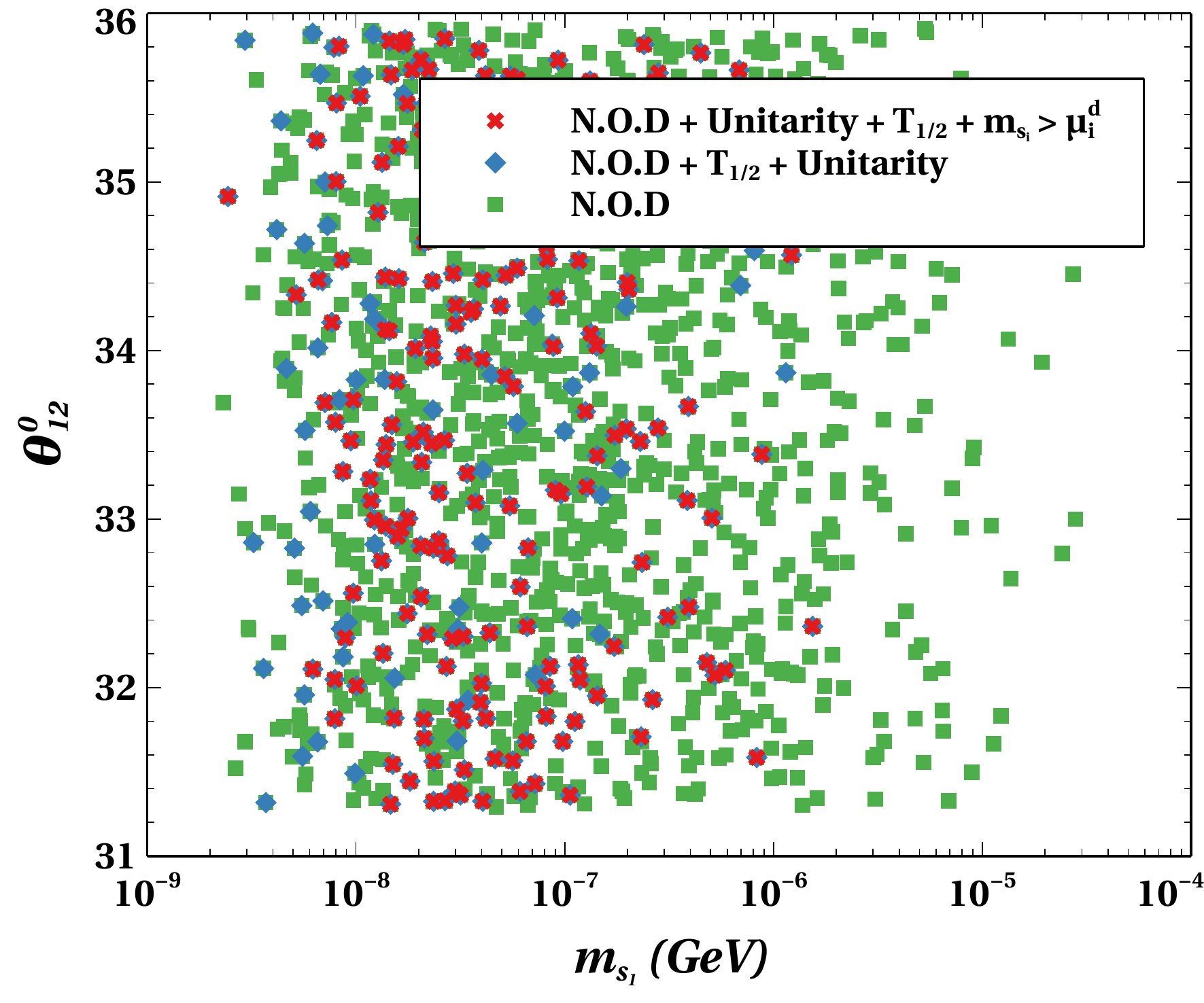}
\caption{Scatter plots in $\sum m_{\nu_i}~ {\rm (eV)}~ - m_{s_1}$ (GeV) (left panel) and
$\theta_{12} - m_{s_1}$ (GeV) (right panel) planes after satisfying the constraints as 
mentioned in the text. N.O.D represents the constraint from 
neutrino oscillation data.} 
\label{fig1-3gen}
\end{figure}
In the left panel of Fig.\,\ref{fig1-3gen}, we have shown the allowed region in $\sum m_{\nu_i}~ {\rm (eV)}~  - m_{s_1}$ (GeV) plane after satisfying all data
\footnote{We get similar kind of behaviour with the other two light sterile neutrinos $s_{2,3}$. Moreover, $s_{2,3}$ also get similar kind of mass  and contribute to $0\nu\beta\beta$ in equal strength.}, where $m_{s_1}$ is the physical mass of the lightest sterile neutrino state.  In the figure, green dots show the range allowed by neutrino oscillation data (N.O.D), blue rhombus points represent the range allowed by $0\nu \beta \beta$ and unitarity along with N.O.D. Finally the red points exhibit the range that is being further constrained by $m_{s_{i}} > \mu_{i}^{d}$. Here, $m_{s_{i}}$ are the physical masses of the sterile state $S_{L}$ and $\mu_{i}^{d}$  are the eigenvalues of $\mu$  and i =1,2,3. In the present work, the model parameters are less constrained from the $0\nu\beta\beta$ decay bound than the unitarity and $m_{s_{i}} > \mu_{i}^{d}$
bounds. 
 %neutrino oscillation data (N.O.D) displayed by green dots, then followed by Unitarity 
%and half lifetime $T_{1/2}$ displayed by blue rhombus, and then finally followed by $m_{s_{i}} > \mu_{i}^{d}$ displayed by red cross points. 
The model parameters range considered in this work give us eV to MeV scale sterile neutrino as seen by the range of $m_{s_1}$-axis.  One interesting thing to note here is that $m_{s_1}$ $\geq 10^{-6}$ GeV is disallowed when we consider both the constraints, unitarity and $0\nu\beta \beta$. This is mainly due to unitarity bound since this bound mostly depends on the ratio of $M_{D}/M_{S}$. Therefore, when $(M_{D}/M_{S})^{2} < 10^{-2}$, those points satisfy the unitarity constraints which is $\pm 10^{-2}$ variation around the unit matrix (see Sec.~\ref{s:constraint}). The disallowed points correspond to higher ratio, {\it i.e.}, $(M_{D}/M_{S})^{2} > 10^{-2}$ and those points represent
lower values of the elements of $\mu$ matrix in order to satisfy the N.O.D which is not covered in Eq. \ref{parameter-range}.   
% to satisfy the N.O.D and hence is disallowed by the 
%Unitarity bound. 
%In Eq.~\ref{e:mass}, one can see that sterile neutrino masses depend on the square of the element of $M_S$ matrix.
%Therefore, after unitarity bound we are getting lower values of the sterile neutrino mass and higher values are disallowed.
This also implies that the elements of $M_{D}$ and $M_{S}$ are of the same order for the disallowed points and more likely to have higher $M_{S}$ values.
Finally the red points are obtained when we impose the constraint $m_{s_i} > \mu_{i}^{d}$. 
%As the range is chosen for the $\mu$ elements, the eigenvalues will be in the eV scale. 
After imposing this constraint, lower values of $m_{s_1}$ are getting ruled out which are mostly in eV scale. In the right panel of this figure we have shown the variation of the solar mixing angle $\theta_{12}$ with the lightest sterile neutrino mass. We can see that the whole allowed range of $\theta_{12}$ from oscillation experiments is in agreement with all the constraints.      
\begin{figure}[h!]
\centering
\includegraphics[angle=0,height=7cm,width=8cm]{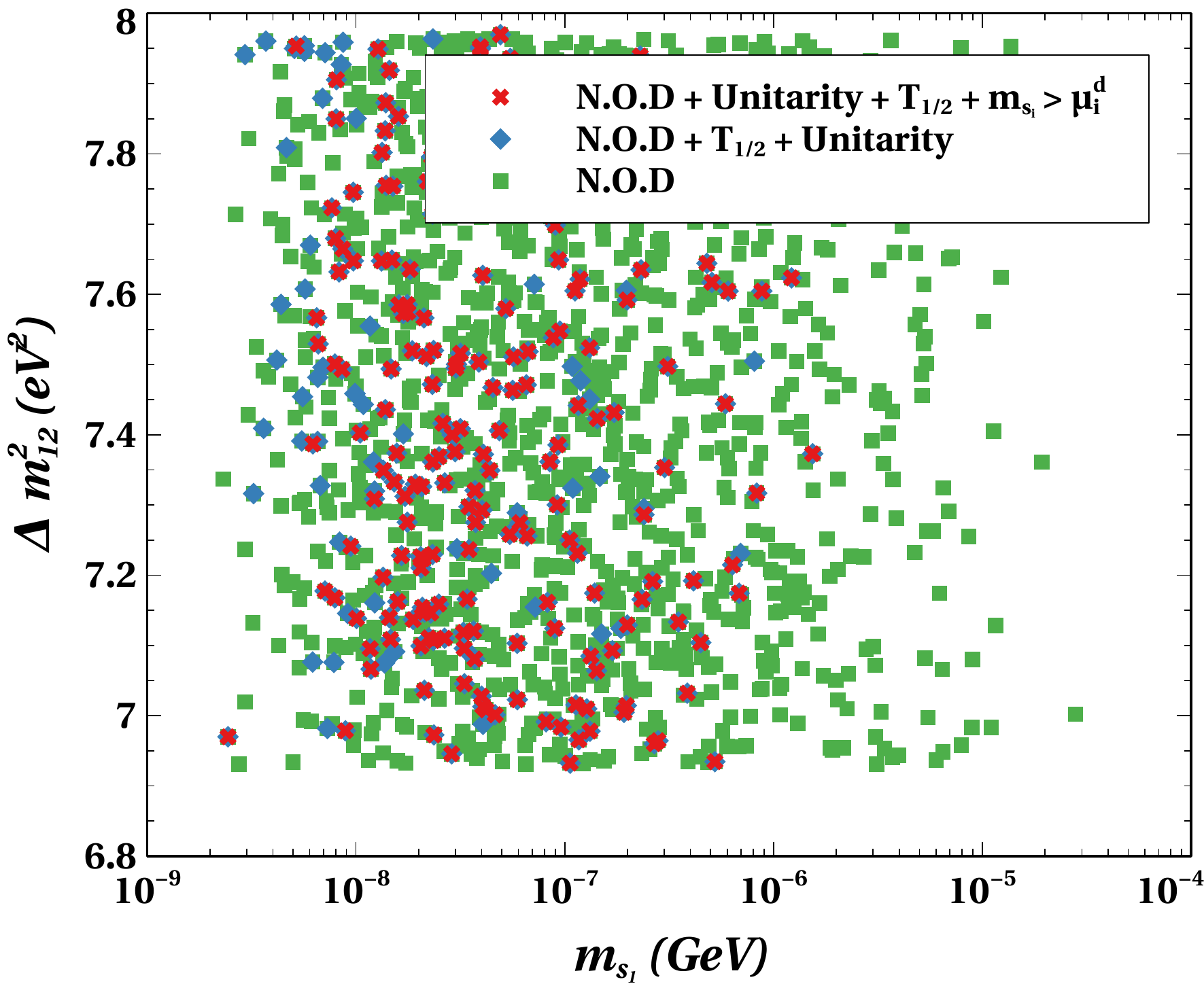}
\includegraphics[angle=0,height=7cm,width=8cm]{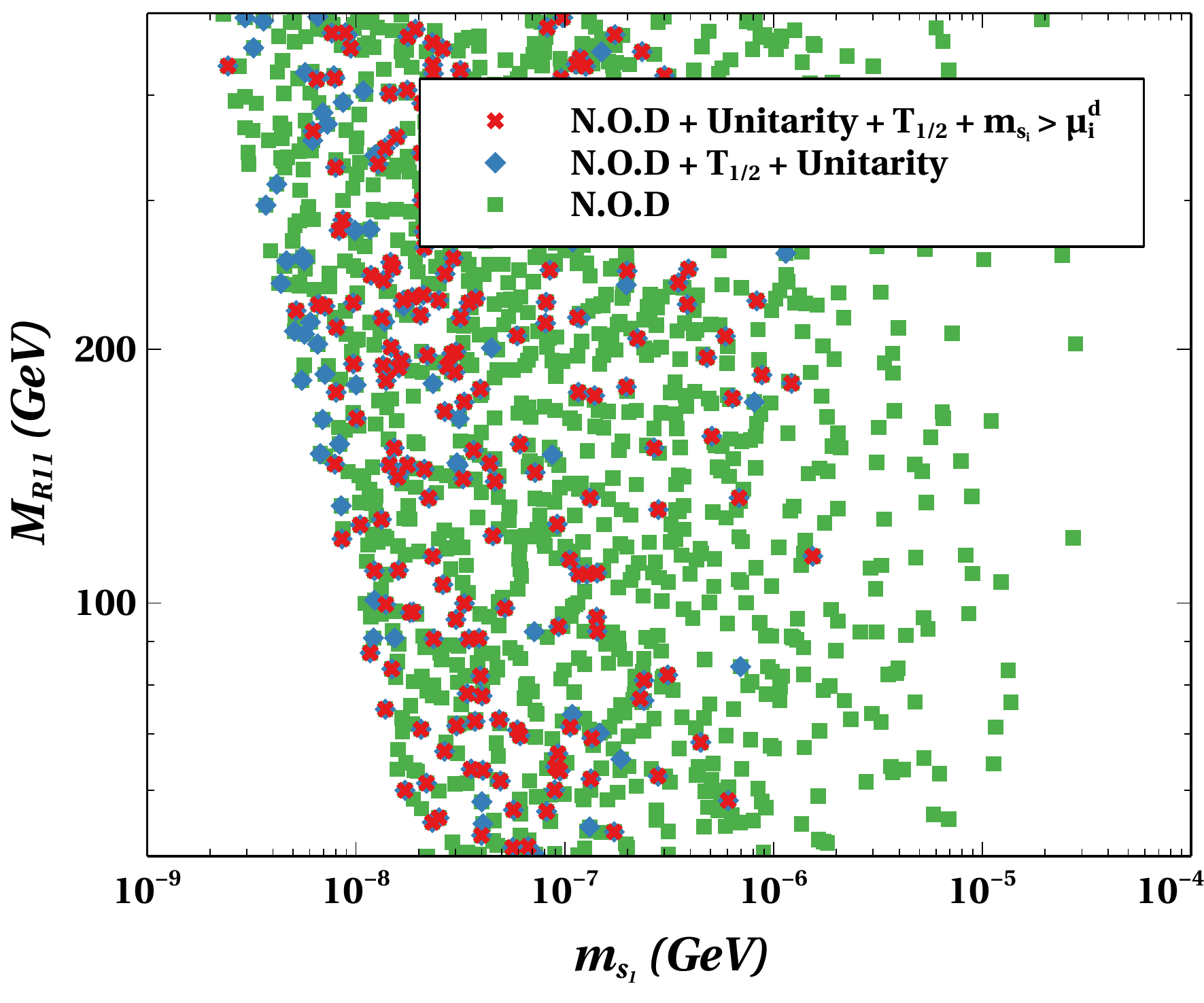}
\caption{Scatter plots in $\Delta m^{2}_{12}~{\rm (eV^{2})} - m_{s_1}$ (GeV)  (left panel) and
$M_{R\,11} - m_{s_1}$ (both are in GeV) (right panel) planes after satisfying the constraints as 
mentioned in the text.}
\label{fig2-3gen}
\end{figure}

In the left panel of Fig.\,\ref{fig2-3gen}, we have shown the scatter plot 
in $\Delta m^{2}_{12}~{\rm (eV^{2})} - m_{s_1}$ (GeV) plane after 
satisfying the constraints as mentioned in the legend of the figure. Here 
also the whole allowed range of $\Delta m^{2}_{12}$ ${\rm (eV^{2})}$ 
from oscillation experiments satisfy all constraints. On the other hand in 
the right panel, we have shown the scatter plot in the 
$M_{R\,11} - m_{s_1}$ plane where $M_{R\,11}$ is the eigenvalue of $m_{n}$
matrix because we have considered here $M_{R}$ as diagonal matrix. 
%In the both panels green points are obtained after satisfying neutrino oscillation data, blue points are obtained after imposing N.O.D + Unitarity + $T_{1/2}$ constraints together and finally the red points survive after imposing the one more constrain which is $m_{s_i} > \mu_{i}^{d}$ (i = 1,2,3).    

\begin{figure}[h!]
\centering
\includegraphics[angle=0,height=7cm,width=8cm]{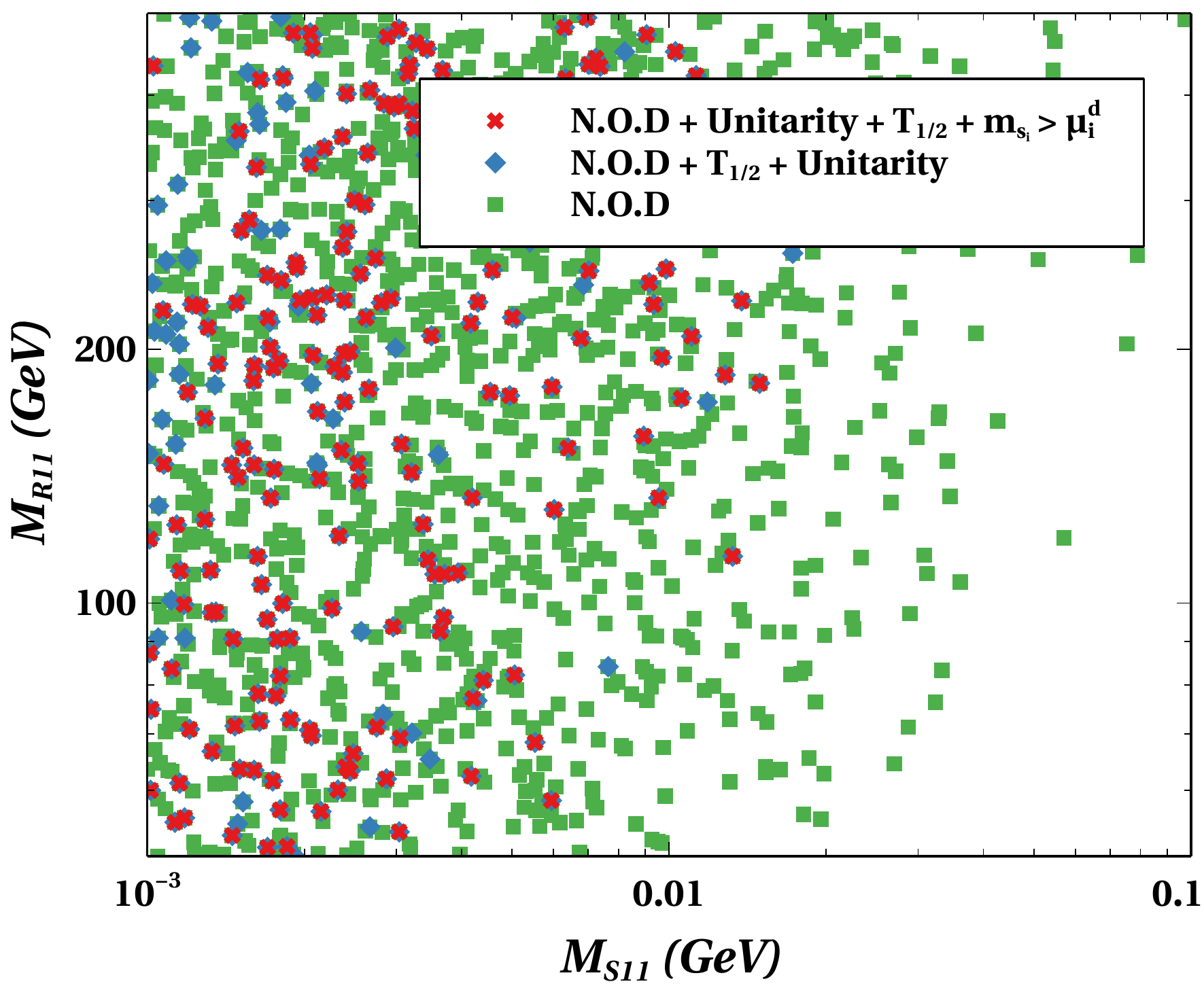}
\includegraphics[angle=0,height=7cm,width=8cm]{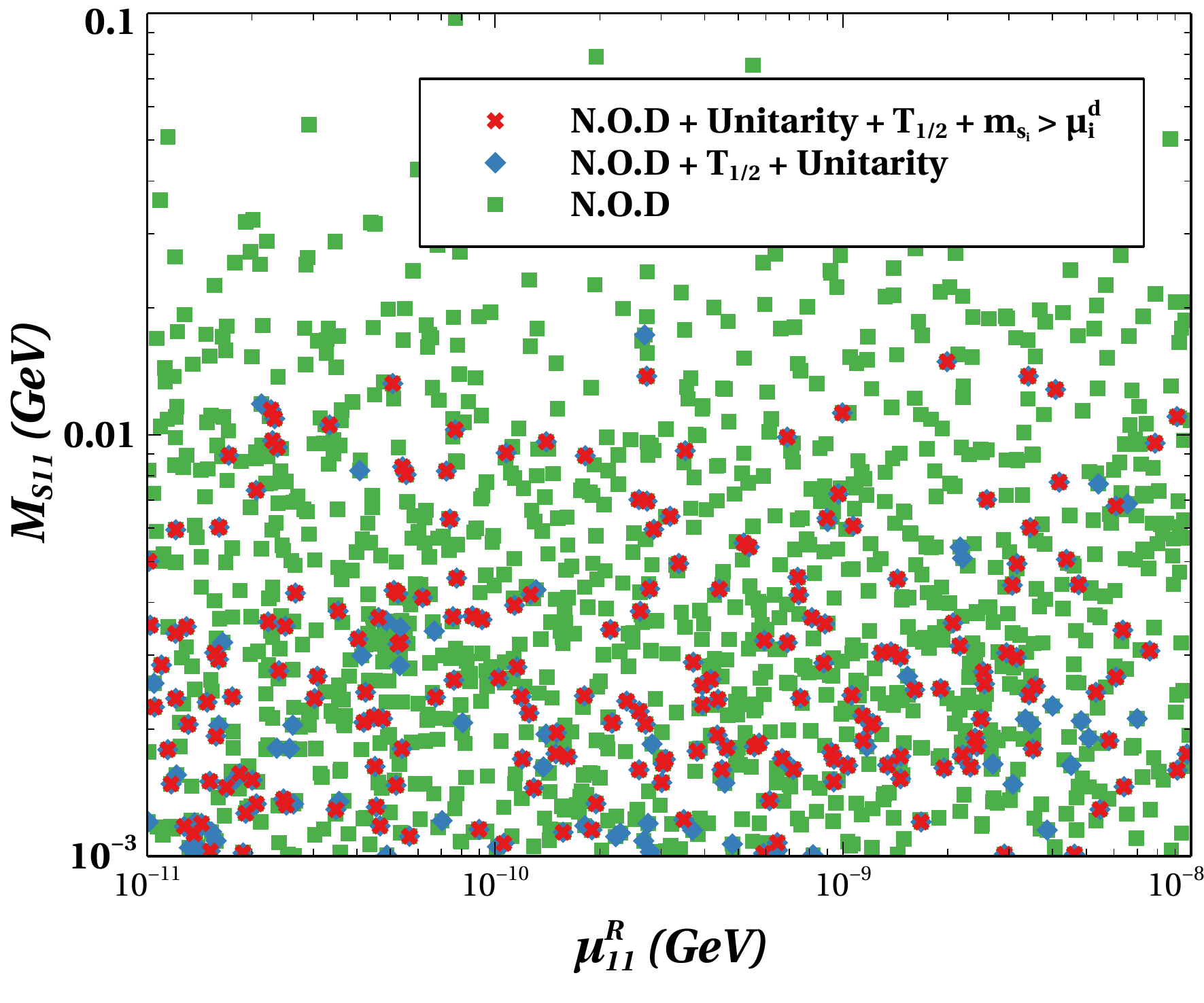}
\caption{Scatter plots in $M_{R\,11} - M_{S\,11}$ (both in GeV) (left panel) and
$M_{S\,11} - \mu^R_{11}$ (both in GeV) (right panel) planes after satisfying the constraints as 
mentioned in the text.}
\label{fig3-3gen}
\end{figure}

Left panel and right panel of Fig.\,\ref{fig3-3gen} shows the variation in $M_{R\,11}$ as function of $M_{S11}$ and $M_{S11}$ as function of $\mu^R_{11}$ respectively. Here, the superscript $R$ implies real elements. %In the LP and RP, we have varied $M_{S\,11}$ \textcolor{red}{(in GeV)} along the x-axis and y-axis, respectively. 
In both cases $M_{S11} > 0.02$ GeV are not allowed by the unitarity constraint as mentioned earlier. The conditions coming from both unitarity and seesaw approximation can be respectively manifested as the upper and lower bound on the light sterile $m_{s}$. 
\begin{figure}[h!]
\centering
\includegraphics[angle=0,height=7cm,width=8cm]{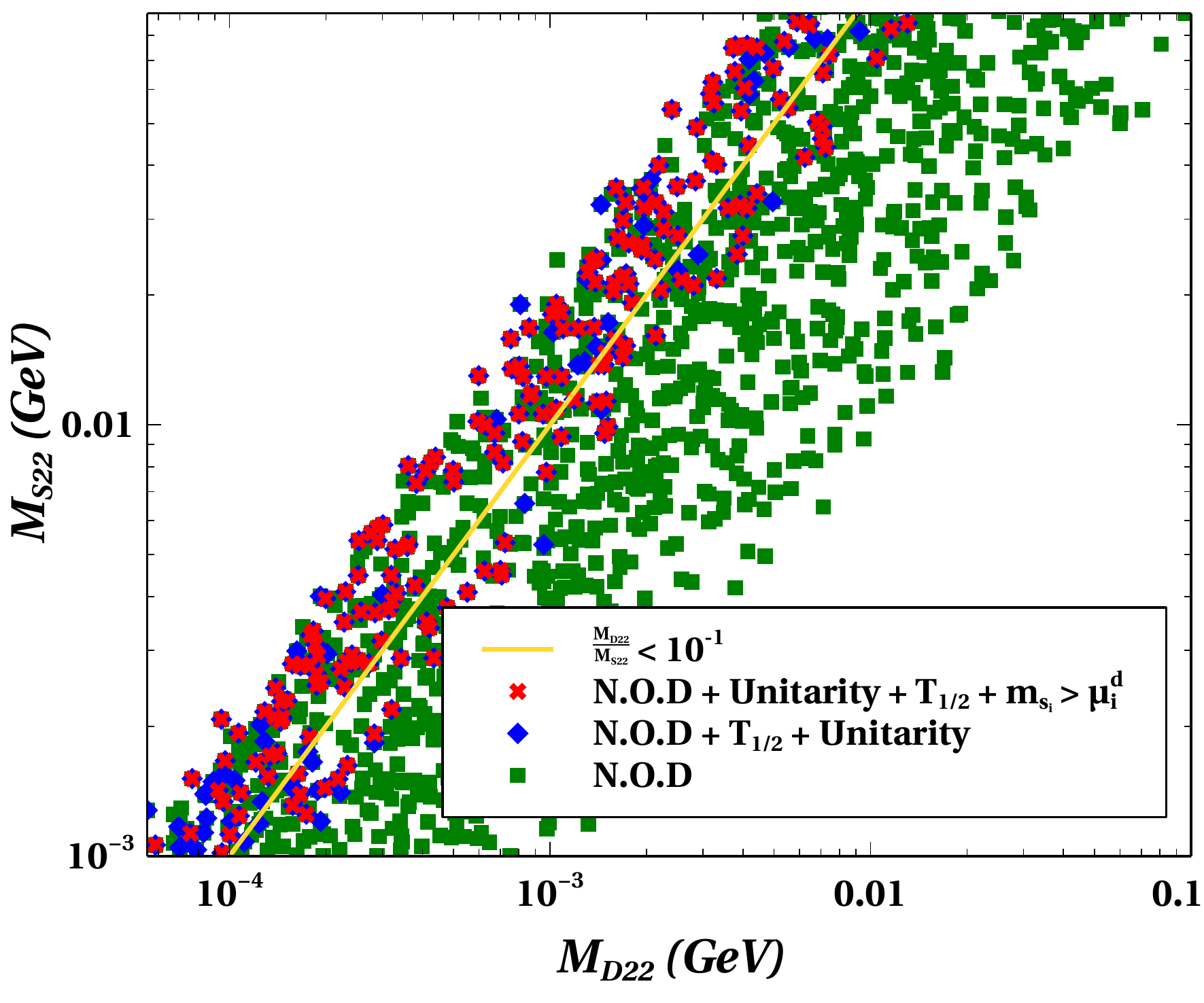}
\includegraphics[angle=0,height=7cm,width=8cm]{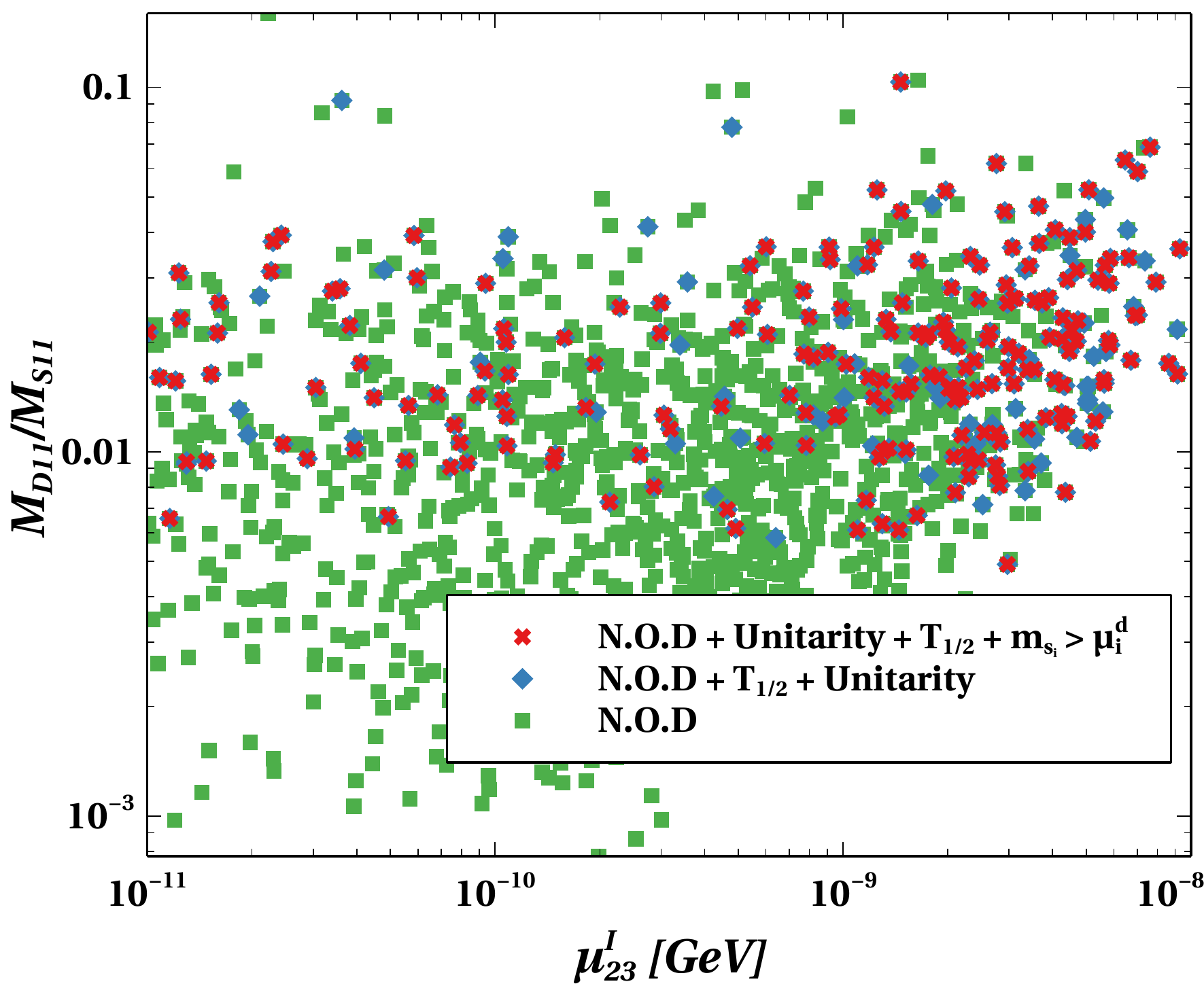}
\caption{Scatter plots in $M_{D22} - M_{S22}$ (both in GeV) (left panel) and
$\mu^{I}_{23}~{\rm(GeV)}~ - \frac{M_{D11}}{M_{S11}}$ (right panel) planes after satisfying the constraints as 
mentioned in the text.} 
\label{fig4-3gen}
\end{figure}

In the left and right panels of Fig.\,\ref{fig4-3gen}, we have shown the scatter plot in  plane $M_{D 22} - M_{S 22}$  and $\mu^{I}_{23}~{\rm(GeV)}~ - \frac{M_{D 11}}{M_{S 11}}$ plane respectively. Here, the superscript $I$ denotes the imaginary part. In the left panel we can see that most of the points which satisfy oscillation data are below the yellow line which  corresponds to $\frac{M_{D 22}}{M_{S 22}} = 10^{-1}$. As we have discussed, the unitarity bound (variation of $\pm 10^{-2}$ around unit matrix) mostly depends on the ratio $\frac{M_{D ii}}{M_{S ii}}$ ($i = 1, 2, 3$) and we can roughly say if $\frac{M_{D ii}}{M_{S ii}} < 10^{-1}$, then it can pass the unitarity bounds. One interesting thing to note here is that there exist a sharp correlation among the $M_{D 22}$ and $M_{S 22}$ parameters (which is valid for other elements of $M_{D}$ and $M_{S}$ matrices also). This is because, approximately $(M_{D}/M_{S})^{2} \mu \sim 10^{-11}$ GeV and we have taken $\mu < 10^{-8} $ GeV, so $M_{D22}$ and $M_{S22}$ can not take arbitrary values which corresponds to significant difference in their magnitudes, otherwise neutrino mass data will not be satisfied. On the other hand in the right panel of Fig.\,\ref{fig4-3gen}, we can see that after imposing all the constraints we get the points which are more prone to have higher values of $\mu^{I}_{23}$. This is because when we impose the unitarity constraint which corresponds to  $\frac{M_{D\, ii}}{M_{S \,ii}} < 10^{-1} $ ($i = 1, 2, 3$) and from the order of magnitude estimation of neutrino mass we obtain $\mu \gtrapprox 10^{-9}$. This is clearly reflected in the right panel of the figure because the points are more dense in that region where $\mu^{I}_{23} \gtrapprox 10^{-9}$ compared to the rest of the region.

\section{Conclusion}
\label{s:conclusion}

In this work, we consider two theory frameworks with sterile neutrinos - a) Left-Right Symmetric Zee, and b) Extended Seesaw model, which successfully explain the light neutrino masses and their mixings. Both of the models can accommodate sterile neutrinos with their masses being free parameters varying over a wide range. We particularly focus on relatively lighter mass range $\sim $ eV to MeV, and explore the contribution to $0\nu \beta\beta$ process. The Left-Right Symmetric Zee model represents a scenario where the masses of the sterile neutrinos are generated at the one loop level. They are  directly dependent on the right-handed Yukawa coupling $\lambda_R$ and the masses and mixings of the charged Higgs bosons. For large values of $\lambda_R$ (close to the perturbative limit) the sterile neutrino masses always remain well below the MeV scale. This presents a unique scenario where the three light sterile neutrinos can have significant contribution to $0\nu \beta\beta$ process.  We find that the half-life of this process crucially depends on three parameters - lightest neutrino mass $m_{\nu_1}$, Dirac CP phase $\delta_{CP}$, and the $W_L-W_R$ mixing angle $\theta_{LR}$. In our analysis, we consider the cases with maximal and minimal mixings among the charged Higgs bosons and also consider both the upper and lower values of the NMEs for ${}^{76}$Ge and ${}^{136}$Xe nuclei.  The scenario with minimal mixing of the Higgs bosons and maximum values of the NMEs produces the most stringent bound on the model. The calculated half-life for both ${}^{76}$Ge and ${}^{136}$Xe nuclei  decreases drastically with an increase in $m_{\nu_1}$ or $\theta_{LR}$. This is due to the dominant contributions coming from the $\lambda$ and $\eta$ diagrams as $m_{\nu_1}$ and/or $\theta_{LR}$ are increased. This allows us to put quite stringent bounds on  both these parameters. For ${}^{76}$Ge nucleus, the lightest neutrino mass should be less than $10^{-7}$ eV for $\theta_{LR} \sim 10^{-4}$ while for a lightest neutrino mass of around $10^{-3}$ eV the value of $\theta_{LR} \lesssim 10^{-8}$, where we consider a normal hierarchy among the active neutrino states. The bounds on ${}^{136}$Xe nucleus are even more stringent with $m_{\nu_1} \lesssim 10^{-8}$ eV for $\theta_{LR} \sim 10^{-4}$ and $\theta_{LR} \lesssim 10^{-8}$ for $m_{\nu_1} \sim 10^{-4}$ eV. Thus we can significantly constrain the model parameters in this case from the $0 \nu \beta\beta$ studies.

For the Extended seesaw, we first consider one-generation scenario where in addition to one SM neutrino two sterile neutrinos are also present. Among the two sterile neutrinos one of them is very heavy with mass of $10^5$ GeV leading to a negligible contribution in $0\nu \beta\beta$ process. The other sterile neutrino has a mass varying in between eV to MeV and this contributes significantly to the above mentioned process. We analyse a number of constraints on the model parameters, arising from $0\nu \beta\beta$, reactor anti-neutrino experiment Daya Bay, non-unitarity constraint on the mixing matrix, as well as, theory-constraints. 
We further extend this simplistic one-generation analysis to higher generation with three active neutrinos and six sterile neutrinos for which we satisfy the neutrino oscillation data.  We present a number of correlations between the mass of the lightest sterile neutrino, model parameters and several neutrino oscillation parameters. In three-generation case, the mass of the heavy sterile states ($N$s) have been varied from 50 to 500 GeV. These sates give negligible contributions to $0\nu\beta\beta$ process, while the other three relatively light sterile states ($S_{L}$s) give substantial contributions.  With the considered parameter range we obtain the upper bound on the mass of the lightest sterile neutrino $S_{1}$ as $10^{-6}$ GeV after imposing constraints from non-unitarity, $0\nu\beta\beta$ and others. The non-unitarity of PMNS matrix has direct impact on the mixing elements; in Extended seesaw, non-unitarity can be governed by the ratio of the bilinear mass term between active neutrino-light sterile states ($M_{D}$) to the corresponding bilinear mass terms among the sterile states ($M_{S}$). The upper bound on the ratio $M_{D}/M_{S}$ is $\sim 10^{-1}$. We choose the effective neutrino mass scale ($\mu M_{D}^{2}/M_{S}^{2}$) to be of the order of $10^{-3}$ eV and we conclude in this scenario the lower bound on the elements of the complex symmetric matrix $\mu$ to be of the order of $10^{-9}$ GeV. Another important constraint in our scenario is $m_{s} > \mu$ below which the seesaw approximation ceases to be valid. It evidently gives lower bound on $m_{s_{i}}$; i.e.,  $\mu_{ii} > 10^{-9}$ GeV implies $m_{s_{i}} > $ 1 eV. The masses of the other two sterile neutrinos vary as $m_{s_{2,3}} \sim \mathcal{O}(1-10)$ eV. %We can eventually have lower values of $m_{s}$ by taking relatively lower values of $\mu_{ii}$ and higher values of $M_{R}$  as $m_{s} \sim M_{S}^{2}/M_{R}$, $M_{R}$ being the mass scale of the sterile $N$.

%\newpage
\paragraph*{Acknowledgements\,:} 
Authors acknowledge Ram Lal Awasthi for his contributions at the initial stage of this project. TJ acknowledges the support from Science and Engineering Research Board (SERB), Government of India under the grant reference no. PDF/2020/001053.

%\newpage
%%%%%%%%%%%%%%%%%   References %%%%%%%%%%%%%%%%%%%%%%%%%%%%%%%%%%%%

\bibliographystyle{JHEP}
\bibliography{reference_ndbd_v8}
\end{document}